\documentclass{appolb}
\usepackage{epsfig}
\usepackage{amstext}
\usepackage{amssymb}
\usepackage{times}
\begin{document}

\title{\Large{\bf Onset of Deconfinement in Pb+Pb Collisions at the Cern SPS}
\vspace{0.5cm}
\begin{center}
{\large P.~Seyboth for the NA49 Collaboration }
\end{center}
\vspace{0.2cm}
\begin{center}
{\normalsize Presented at the XLVI Cracow School of Theoretical Physics,
Zakopane, Poland, May 27 - June 5, 2006 }
\end{center}
\vspace{0.2cm}
\begin{center}
{\normalsize (to be published in Acta Physica Polonica B) }
\end{center}
}

\maketitle

\vspace{2cm}
\begin{abstract}

The NA49 fixed-target experiment studied high 
energy--density matter produced in nucleus--nucleus
reactions at the CERN SPS. In central Pb+Pb collisions at 158$A$ GeV the
energy density at the early stage substantially exceeds the threshold for quark
deconfinement predicted by lattice QCD. The produced matter shows strong
transverse and longitudinal flow. Ratios of yields of produced particles are 
approximately consistent with
statistical equilibration. An energy scan through the
SPS range revealed structure in the energy dependence of $\pi$ and K yields
as well as of the inverse slopes of transverse mass distributions. These features suggest
that a deconfined phase starts to be produced at around 30$A$ GeV in
central Pb+Pb collisions. The analysis of fluctuations and correlations has not
yet provided evidence for the predicted critical point of QCD. 

\end{abstract}

\newpage

\noindent
C.~Alt$^{9}$, T.~Anticic$^{23}$, B.~Baatar$^{8}$,D.~Barna$^{4}$,
J.~Bartke$^{6}$, L.~Betev$^{10}$, H.~Bia{\l}\-kowska$^{20}$,
C.~Blume$^{9}$,  B.~Boimska$^{20}$, M.~Botje$^{1}$
% ...

\vspace{0.5cm}
\noindent
$^{1}$NIKHEF, Amsterdam, Netherlands. \\
% ...
$^{4}$KFKI Research Institute for Particle and Nuclear Physics, Budapest, Hungary.\\
% ...
$^{6}$Institute of Nuclear Physics, Cracow, Poland.\\
% ...
$^{8}$Joint Institute for Nuclear Research, Dubna, Russia.\\
$^{9}$Fachbereich Physik der Universit\"{a}t, Frankfurt, Germany.\\
$^{10}$CERN, Geneva, Switzerland.\\
% ...
$^{20}$Institute for Nuclear Studies, Warsaw, Poland.\\
% ...
$^{23}$Rudjer Boskovic Institute, Zagreb, Croatia.\\

\newpage

\vspace{-0.3cm}

\section{Introduction}\label{introduction}

Qualitative considerations based on the finite size of hadrons \cite{col75}
as well as quantum chromodynamics (QCD) calculations on the lattice
\cite{kar02,fod04} predict that at sufficiently high energy density strongly
interacting matter will transform into a state of quasi-free quarks and gluons,
the quark gluon plasma (QGP). Theoretical investigations also found \cite{step98} 
that this phase
transition is of first order for finite quark masses and large non--zero baryon
density. The phase boundary is predicted to end in a critical point and turn
into a rapid crossover as the net baryon density decreases.

\begin{figure}[hbt]
\begin{center}
\epsfig{file=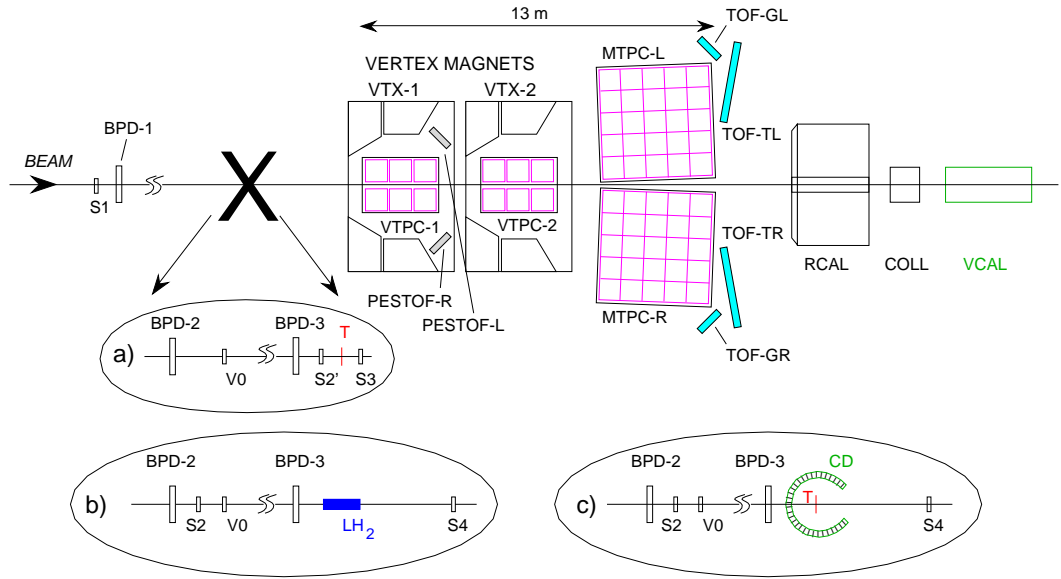,width=14.0cm}
\end{center}
\caption{Schematic layout of the NA49 experiment at the CERN SPS showing
beam detectors, superconducting dipole magnets, time projection
chambers (VTPC, MTPC), time-of-flight arrays (TOF) and calorimeters (RCAL, VCAL).
A thin solid target T is used for A+A collisions (a), which is surrounded by
a detector of slow protons (CD) for p+A collisions (c). A liquid H$_2$ target is
employed for p+p collisions (b)}
\label{fig_na49}
\end{figure}

The initial stage of high energy collisions of large nuclei provides the best
environment to produce the deconfined phase of matter in the laboratory \cite{qm04,qgpsign}.
Lead ions first became available at the CERN SPS in 1994. In the first publication
from this programme \cite{alb95} NA49 demonstrated
that in central Pb+Pb collisions at top SPS energy the initial energy density
exceeds the critical value of $\approx$~1~GeV/fm$^3$. The SPS experiments found
that the reactions produced an explosively expanding fireball.
Moreover, originally proposed signatures of the QGP,
i.e. $J/\Psi$ suppression, strangeness enhancement, and possibly
thermal photons and dileptons were observed \cite{cern2000}.
However, these signatures are not specific for deconfinement.
The NA49 collaboration therefore performed an energy scan from
20~-~158$A$ GeV in order to search for structure in the energy dependence
of hadron production characteristics which could indicate the onset
of deconfinement \cite{GaGo99}. The measurements indeed suggest structure
around 30$A$ GeV \cite{na49_qm2004} which will be discussed below. NA49 also 
searched for fluctuations which might occur if distinct phases coexisted in
the early stage of the reactions or if hadrons froze out close to the 
critical point estimated \cite{fod04} to lie in a region accessible at the SPS.

\section{Experiment NA49 at the CERN SPS}\label{detector}

The main features of the NA49 experiment \cite{na49} located in the H2
beam line of the North Experimental Hall (see fig.~\ref{fig_na49}) are large acceptance
precision tracking ($\Delta p/p^2 \approx(0.3 - 7)\cdot 10^{-4}$(GeV/c)$^{-1}$)
and particle identification in the central and forwards rapidity regions
using time projection chambers (TPCs). Charged particles ($\pi$, K, p, $\bar{\rm p}$)
are identified mostly from the measurement of their energy loss in the
TPC gas (accuracy 3 -- 6 \%). Yields are obtained by fitting a sum of
Gaussian functions for the various particle species to the dE/dx distributions
in small bins of momentum p and transverse momentum $p_T$. At central rapidity the
identification is further
improved by measurement of the time-of-flight (resolution 60 ps) to arrays of
scintillation counter tiles. Strange particles (K$^0_s$, $\Lambda$, $\Xi$, $\Omega$)
are detected via decay topology and invariant mass measurement.
A forward calorimeter measures the energy of the projectile spectators from which
the impact parameter in A+A collisions and the number of participating nucleons
are deduced. Also reactions of C and Si ions as well as of deuterons were studied.
These beams were produced
by fragmentation of the primary Pb beam and selected by magnetic rigidity and
by specific energy loss in transmission detectors in the beam line. In addition
NA49 pursued an extensive program of proton induced collisions for a study of 
the evolution of particle production from p+p via p+Pb to Pb+Pb reactions.

\begin{figure}[hbt]
\begin{center}
\mbox{
 \parbox{5.0cm}{\hspace{-1.0cm}
  \epsfig{
            file=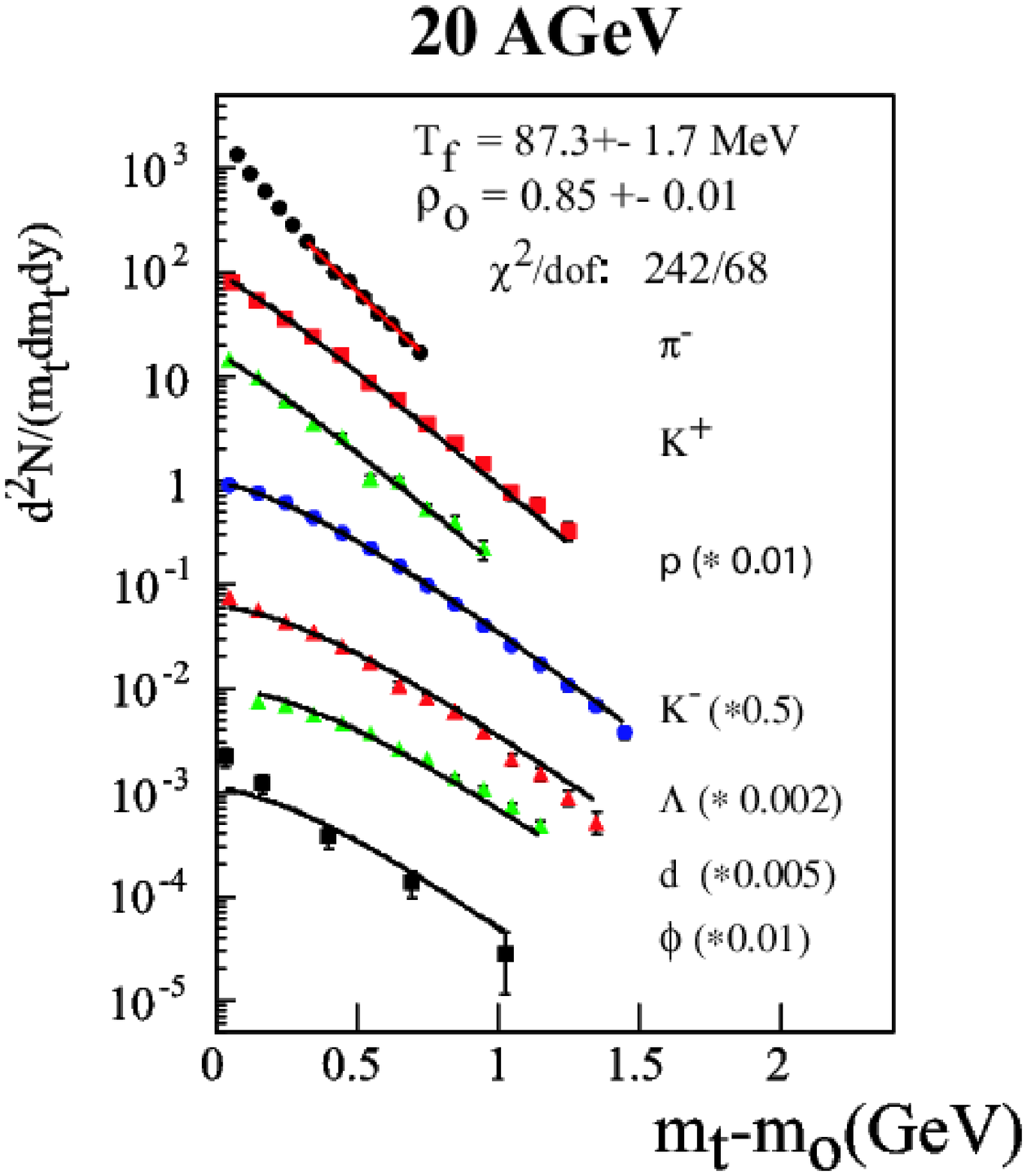,width=5.0cm}}
 \parbox{5.0cm}{\hspace{-1.0cm}
  \epsfig{
            file=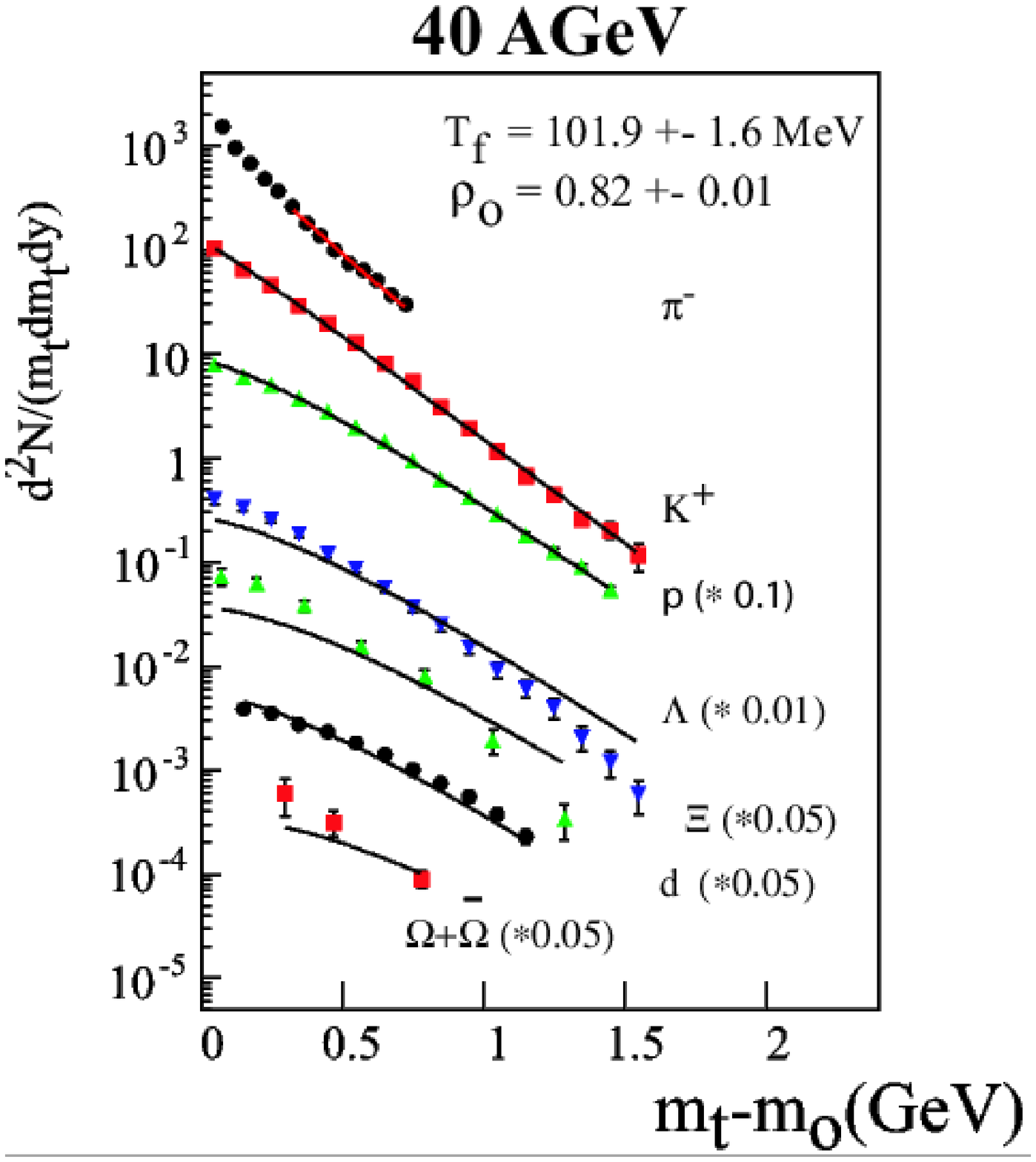,width=5.0cm}}
 \parbox{5.0cm}{\hspace{-1.0cm}
  \epsfig{
            file=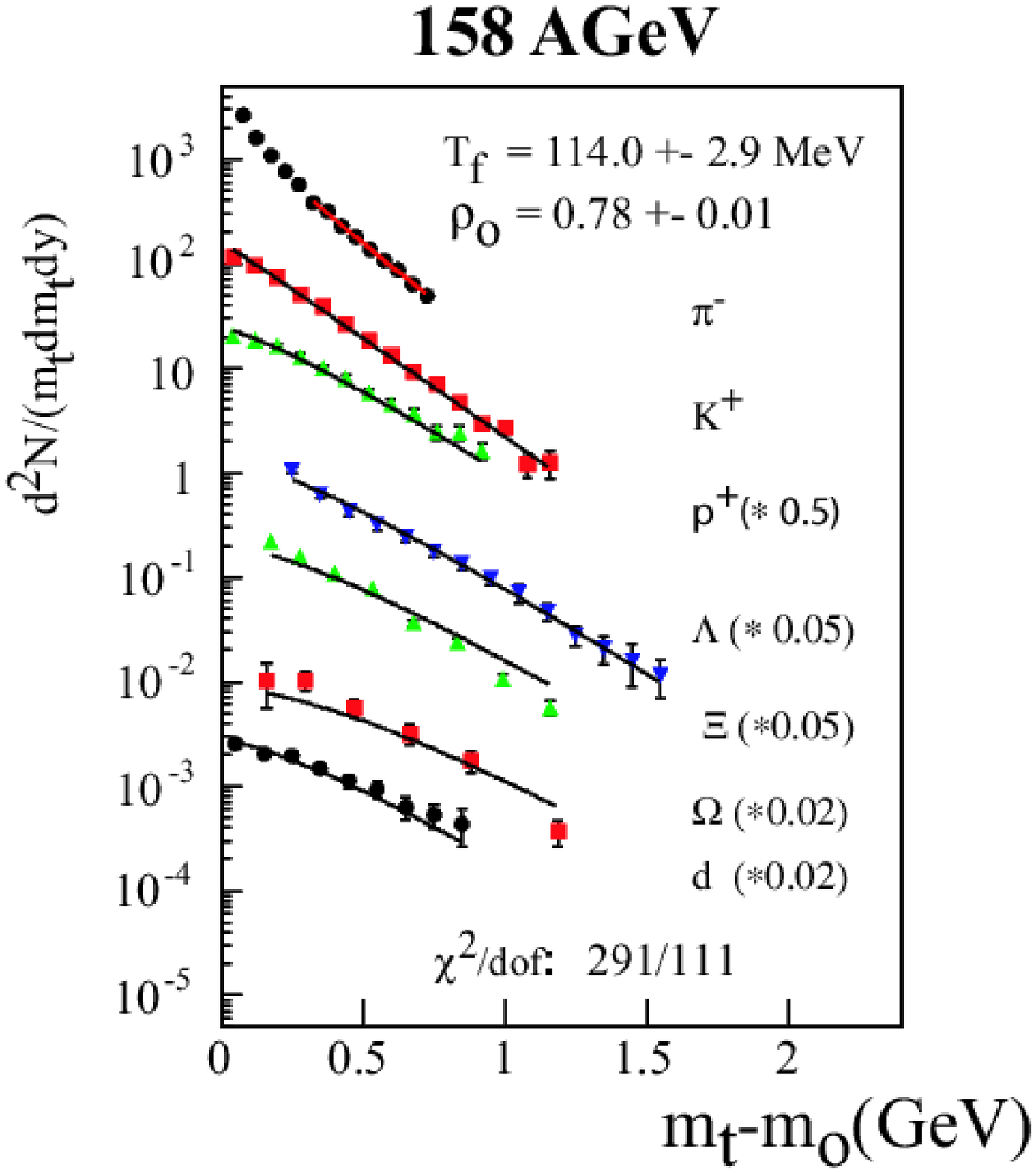,width=5.0cm}}
}
\end{center}
\caption{
Transverse mass $m_t$ spectra at midrapidity in central Pb+Pb collisions 
at 20$A$ (left), 40$A$ (center) and 158$A$ (right) GeV. The curves show 
the result of a blast wave fit \cite{lisa2004} with
parameters: temperature T$_f$ and surface radial flow velocity $\rho_0$.
}
\label{fig1}
\end{figure}

\section{Analysis procedure}\label{analysis}
All events used for physics analysis were required to have a good vertex in the
respective target. Tracks of accepted particles had to fulfill quality requirements
on the number of measured points and the distance of closest approach to the
event vertex. Particle yields have been corrected for acceptance, reconstruction efficiency
and feed-down contamination from weak decays. Results on particle yields for the upper 3
energies have been published in \cite{Afa02,appe98,na49_omega} where more details on the analysis
procedures may be found.  Results at 20$A$ and 30$A$ GeV are preliminary.

\section{Thermal freezeout parameters from spectra and correlations}\label{thermfo}

Midrapidity invariant yields as a function of the transverse mass $m_t$ are shown in
fig.~\ref{fig1} at 20$A$, 40$A$ and 158$A$ GeV as examples
for the large variety of particle species measured
by NA49. The spectra become progressively flatter with increasing particle
mass, a fact that can be explained by the combined effect of the random thermal momentum
distribution and strong radial flow in the produced matter droplet. The development of
collective flow is generally attributed to the hydrodynamic pressure generated in the
dense early stage \cite{heinz2000}. 
A hydrodynamically based "blast wave" parameterization \cite{lisa2004} indeed
provides a reasonable description of all the spectra with two parameters:
a temperature T$_f\approx$~(90~-~110)~MeV and a radial flow velocity
of $\rho_0 \approx$~0.8~c at the surface. These parameters characterize the thermal/kinetic
freezeout of the fireball. The pion spectrum overshoots the model curve at low $m_t$
due to the feed-down contribution from resonance decays and this region was therefore not
used in the fit. More detailed analyses \cite{anti2004,na49_omega} seem to suggest an earlier 
freezeout of multistrange hyperons at higher temperature and smaller radial flow velocity.

\begin{figure}[hbt]
\mbox{
 \parbox{6.0cm}{
  \epsfig{figure=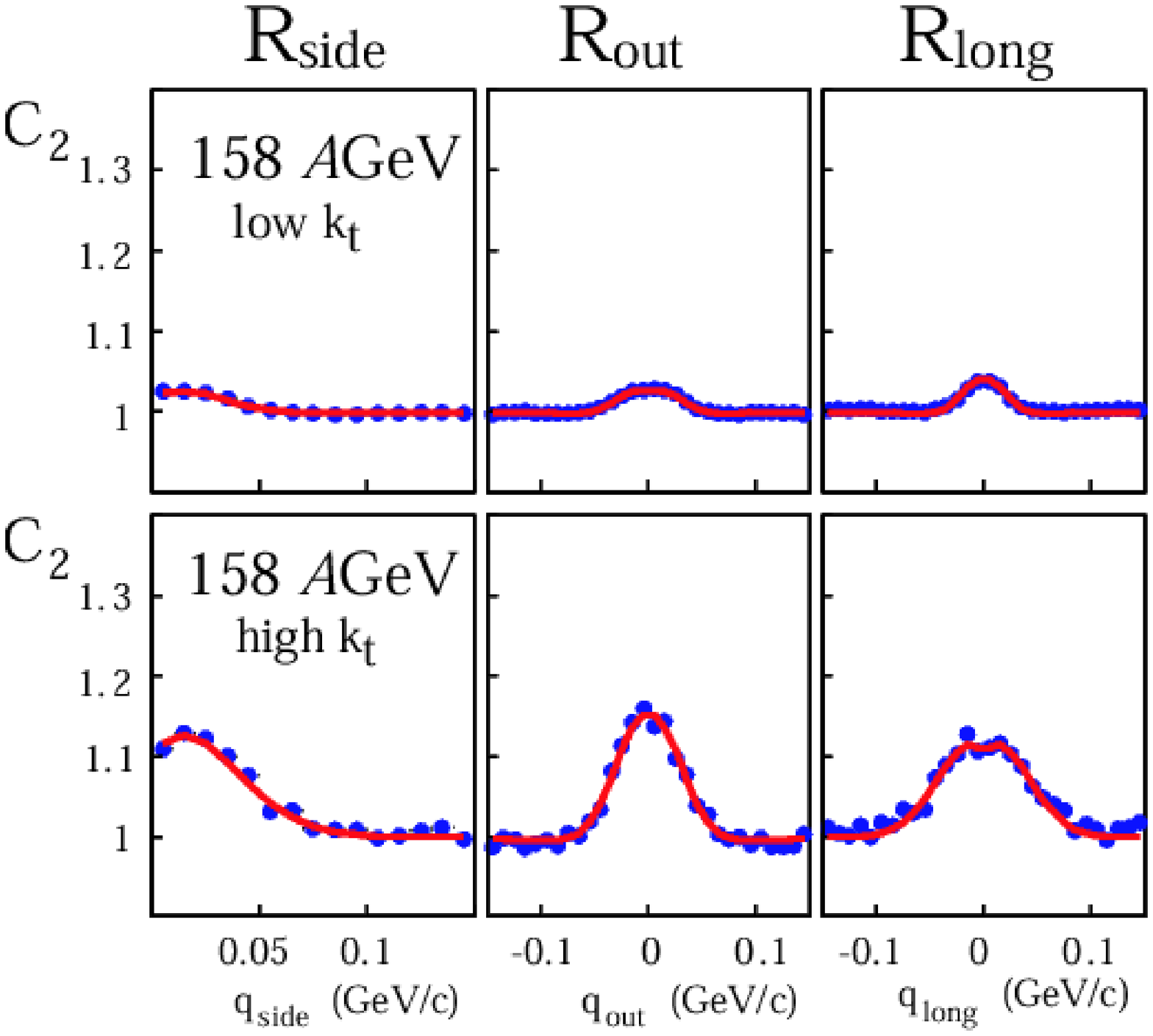,width=6.0cm}}
 \parbox{6.0cm}{
  \epsfig{figure=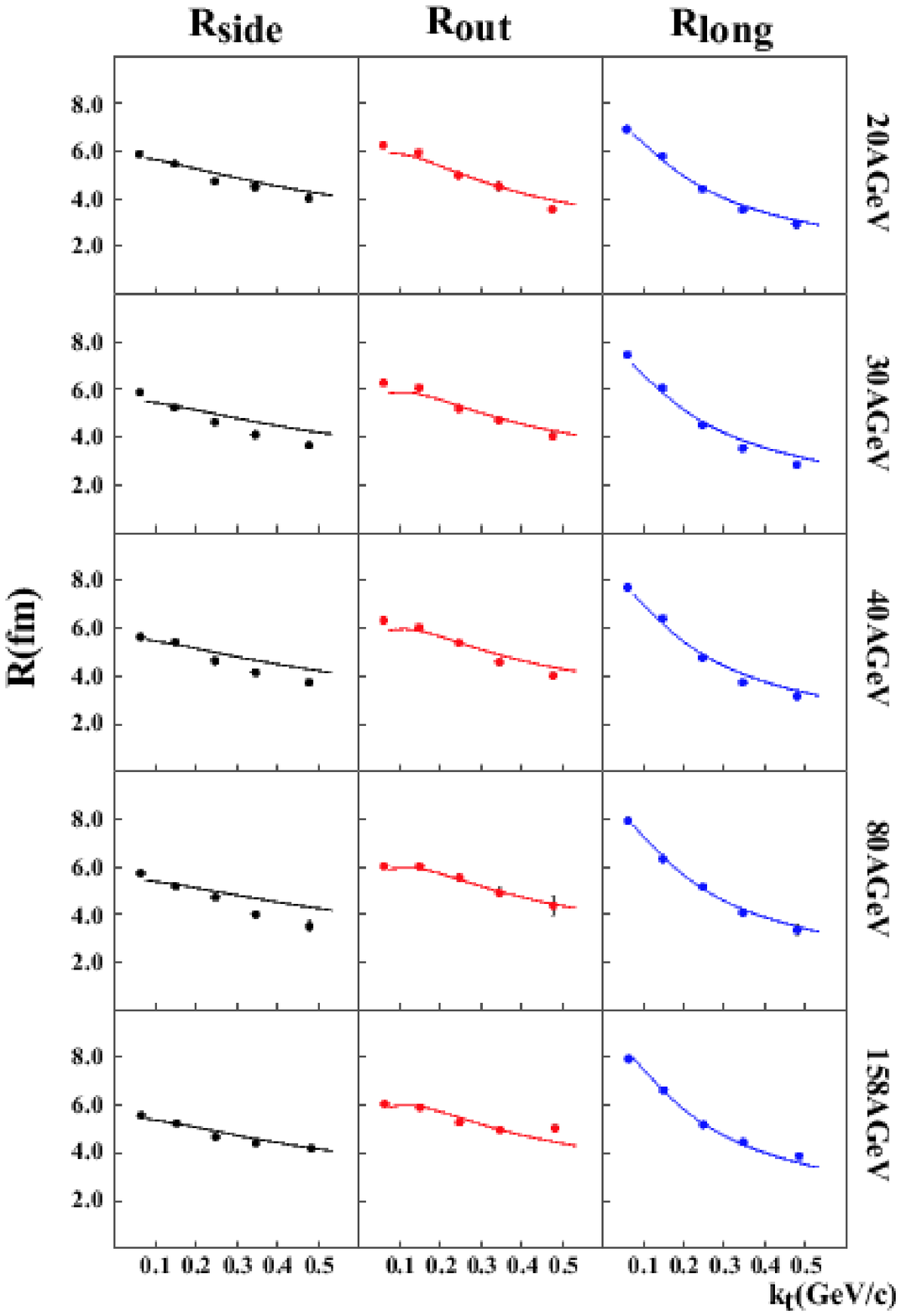,width=6.0cm}}
}
\caption{
Left: Projections of the $\pi^-\pi^-$ correlation function on momentum difference q in the
transverse (side, out) and beam (long) directions for central Pb+Pb collisions at
158$A$ GeV. 
Right: Effective radius parameters R of the pion source versus mean transverse
momentum $k_t$ of the pion pairs in central Pb+Pb collisions at the five SPS energies.
The curves show results of simultaneous fits of the blast wave model to the radius parameters
and the transverse mass $m_t$ spectra of pions and protons. All results refer to midrapidity
pion pairs.
}
\label{be_radii}
\end{figure}

\begin{figure}[hbt]
\mbox{
 \parbox{6.0cm}{
  \epsfig{figure=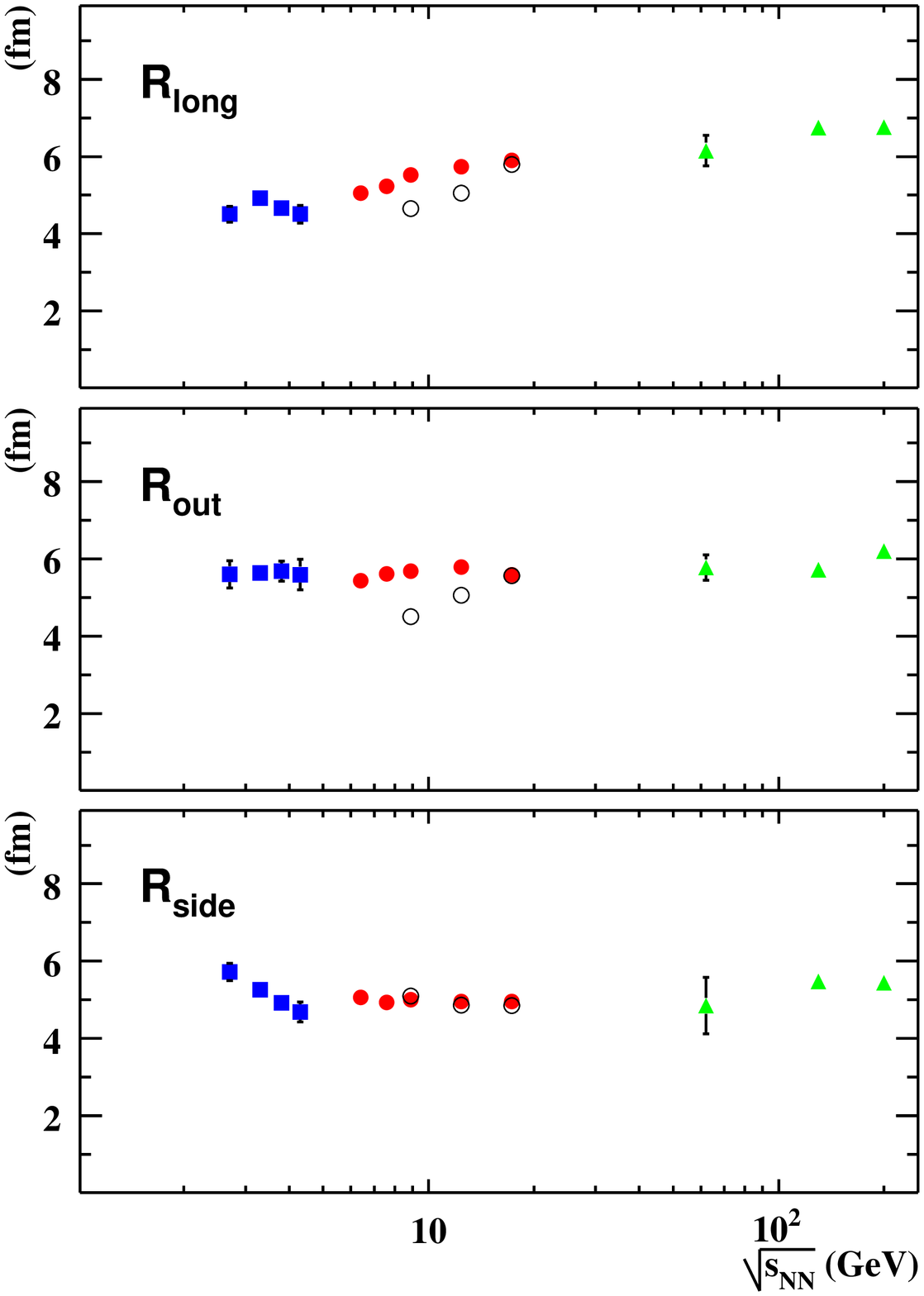,height=8.0cm}}
 \parbox{6.0cm}{
  \epsfig{figure=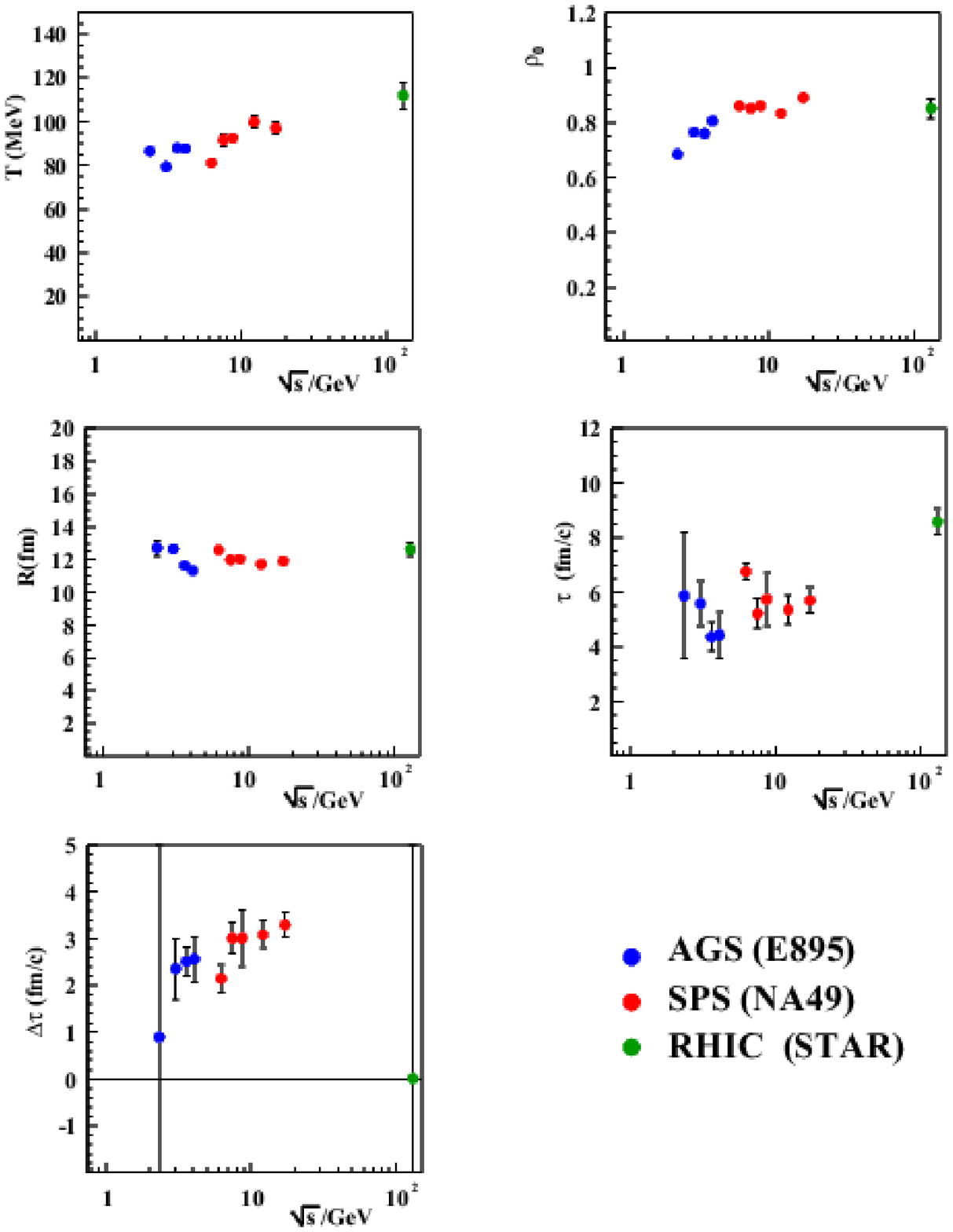,height=8.0cm}}
}
\caption{Left: Energy dependence of radius parameters R of the pion source 
at mean transverse momentum $k_t$= 0.2 GeV/c and midrapidity in central Pb+Pb and
Au+Au collisions. Results from AGS, SPS and RHIC are shown as squares, filled (NA49)
and open (NA45) circles, and triangles respectively.
Right: Freezeout parameters obtained from simultaneous fits of the blast wave model ~\cite{lisa2004} 
to radius parameters from $\pi^-\pi^-$ Bose-Einstein correlations and $m_t$ spectra of
pions and protons at midrapidity in central Pb+Pb (Au+Au) collisions from AGS (blue dots)
through SPS (red dots, NA49 data) to RHIC (green dots) energies.
}
\label{be_edep}
\end{figure}

Collective flow also affects quantum statistics induced correlations of identical particles.
For negative pion pairs the range of these Bose-Einstein (BE) correlations in momentum difference,
which is inversely proportional to the effective pion source size, is reduced by flow. This reduction
gets stronger with increasing transverse momentum (see fig.~\ref{be_radii}~(left)). 
In the NA49 publication \cite{ykp} this
effect together with the $m_t$ spectrum of pions was used for the first time to
demonstrate strong collective radial flow in Pb+Pb collisions, i.e. the correlation of
the radial position of the emission points with the transverse momentum of the produced
particle. This information cannot be obtained from transverse mass spectra alone. 
The fitted effective source sizes, the
so-called radius parameters R, are shown as a function of averaged transverse momentum $k_t$
of the pion pairs for all the SPS energies in fig.~\ref{be_radii}~(right).
The decrease of R with increasing $k_t$ due to radial flow is clearly observed. On the other hand,
as shown in fig.~\ref{be_edep}~(left) the radius parameters show remarkably little energy
dependence over the whole energy range from AGS to RHIC. In particular, there is no indication
of an increase of R$_{\text{out}}$ at the SPS which might be expected from a first order
phase transition. In fact, hydrodynamic calculations with a simple freezeout procedure overpredict
the values of R$_{\text{out}}$ and R$_{\text{long}}$ \cite{heinz2000}. A solution of
this so-called HBT puzzle seems to lie in a more sophisticated treatment of the frezzeout process
\cite{siny_2006}.

The hydrodynamics inspired blast wave parameterisation provides a comprehensive description
of the kinetic freezeout stage of nucleus--nucleus reactions.
The parameterization ~\cite{lisa2004} assumes a uniform pion emission
density in a cylinder of radius R and a radial flow velocity increasing linearly
with radius to a surface maximum of $\rho_0$. Further parameters of the model are the kinetic
freezeout temperature T (mainly determined by the $m_t$ spectra), the lifetime $\tau$ of the
fireball (derived mostly from R$_{\text{long}}$) and the duration $\Delta\tau$ of the
pion emission burst (obtained from the difference of R$_{\text{out}}$ and R$_{\text{side}}$).
Resulting parameter values from simultaneous fits to radii and $m_t$ spectra from AGS to RHIC
energies are plotted versus collision energy in fig.~\ref{be_edep}~(right). The corresponding 
radius parameters at the SPS are shown as curves in fig.~\ref{be_radii}~(right). 
The fit results demonstrate that the matter droplet at freezeout has increased radially
by a factor of two and has attained strong collective radial flow. While frezeout temperature 
and fireball lifetime appear to to increase slowly with energy, the surface radial 
flow velocity seems to saturate at SPS energies.

While radial flow manifest in $m_t$ spectra and BE correlations builds up over the
full lifetime of the fireball, anisotropic flow generated in non-central collisions
is particularly sensitive
to the properties of the produced matter during the early phase of the reaction. 
In the almond shaped interaction region anisotropic pressure gradients and rescattering 
lead to an azimuthally anisotropic momentum distribution. Since the spatial anisotropy
disappears quickly, the momentum anisotropy is generated predominantly at early times
in the reaction. The pressure rise with initial energy density is qualitatively expected
to slow down when deconfinement sets in. Anisotropic flow is quantified by the Fourier
expansion coefficients of the azimuthal angular distribution with respect to the
reaction plane. At midrapidity the dominant effect is observed in the second coefficient v$_2$, 
termed elliptic flow.

\begin{figure}[hbt]
\begin{center}
\mbox{
 \parbox{6.0cm}{
  \epsfig{bbllx=30bp,bblly=50bp,bburx=570bp,bbury=750bp,
          figure=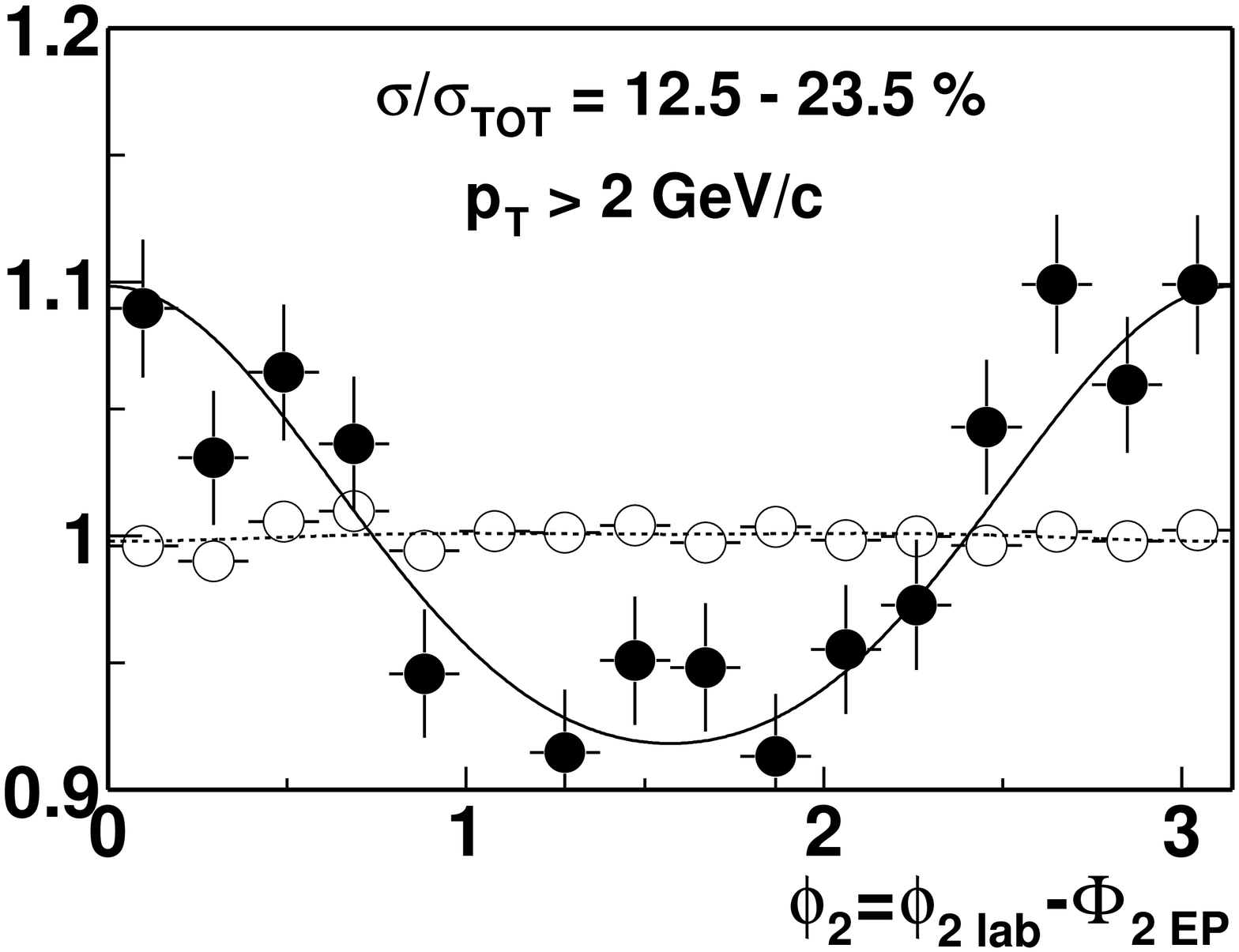,angle=-90,width=6.0cm}}
 \parbox{6.0cm}{
  \epsfig{bbllx=0bp,bblly=0bp,bburx=550bp,bbury=620bp,
          figure=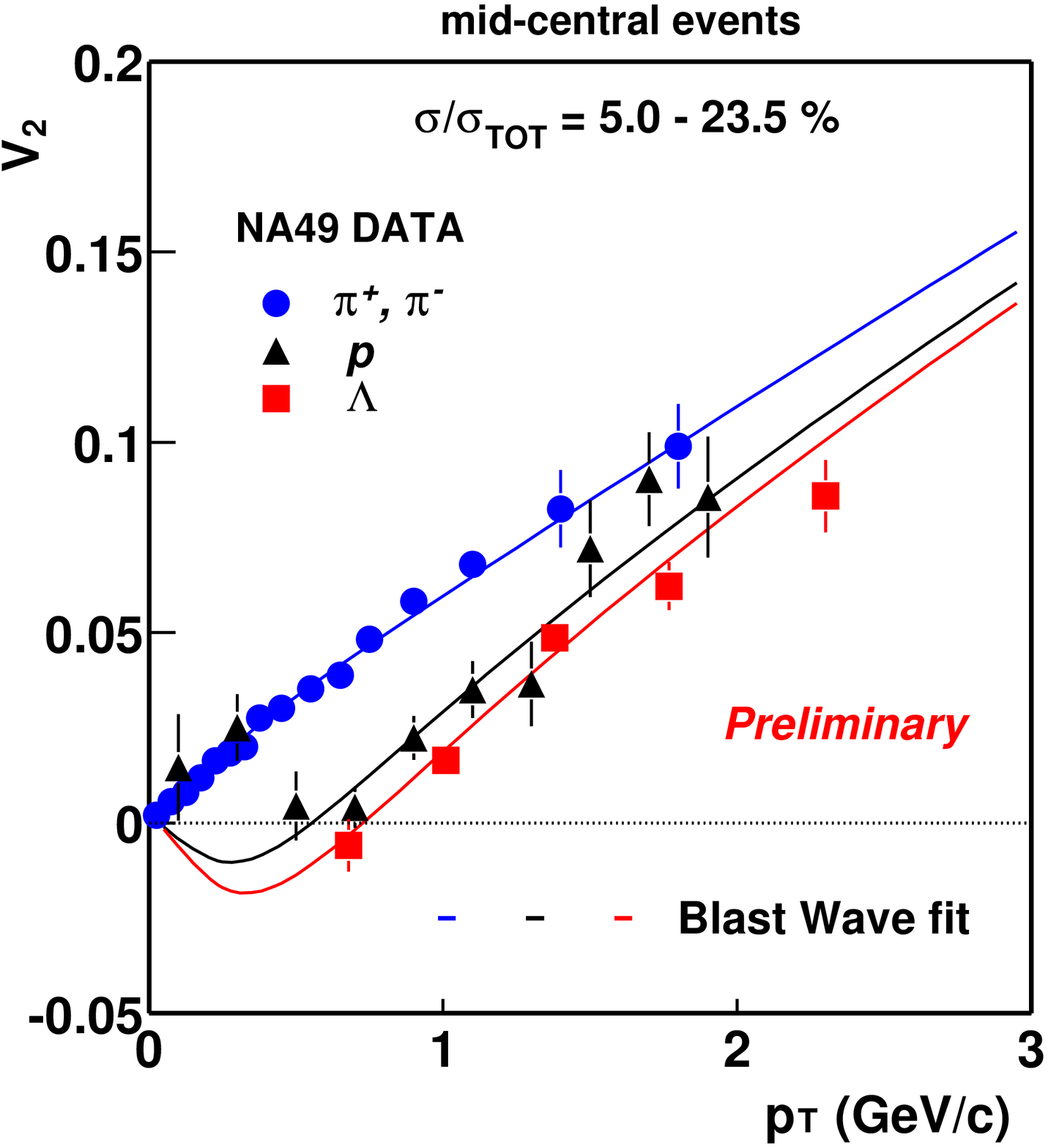,width=6.0cm}}
}
\end{center}
\caption{
Left: Azimuthal distribution of $\Lambda$ hyperons with respect to the reaction
plane as determined from pions. Full dots show data, open dots mixed events.
Right: v$_2$ of $\pi$, p, $\Lambda$ versus transverse momentum $p_T$ at midrapidity 
in Pb+Pb collisions at 158$A$ GeV. Curves show results from the blast wave model
\cite{lisa2004}.
}
\label{lambdav2}
\end{figure}

NA49 measured anisotropic flow of pions and protons in Pb+Pb collisions at 
40$A$ and 158$A$ GeV \cite{flow03} and of $\Lambda$ hyperons \cite{stef05} at 158$A$ GeV. 
An example of the $\Lambda$ azimuthal distribution with respect to the reaction plane 
as determined from pions is displayed in fig.~\ref{lambdav2}~(left). 
The elliptic flow coefficient v$_2$ at midrapidity is plotted versus 
transverse momentum $p_T$ for the different particle species in fig.~\ref{lambdav2}~(right). 
One observes a strong increase with $p_T$ and a clear mass hierarchy as predicted by
hydrodynamic models \cite{heinz2000}. Values of v$_2$ are about 30 \% higher at 
RHIC (not shown) than at SPS. 
Both at the SPS and at RHIC the observed elliptic flow v$_2$ can be reproduced by 
the blast wave model parameterisation \cite{lisa2004} with parameters
consistent with those describing spectra and Bose--Einstein correlations \cite{stef05}. Pure
hydrodynamic model calculations overpredict v$_2$ particularly at SPS energies. However,
coupling of a hadronic rescattering phase to the initial deconfined hydrodynamic phase
results in reasonable description of the measured energy dependence from SPS to RHIC \cite{tean_2001}.

\begin{figure}[hbt]
\begin{center}
\epsfig{file=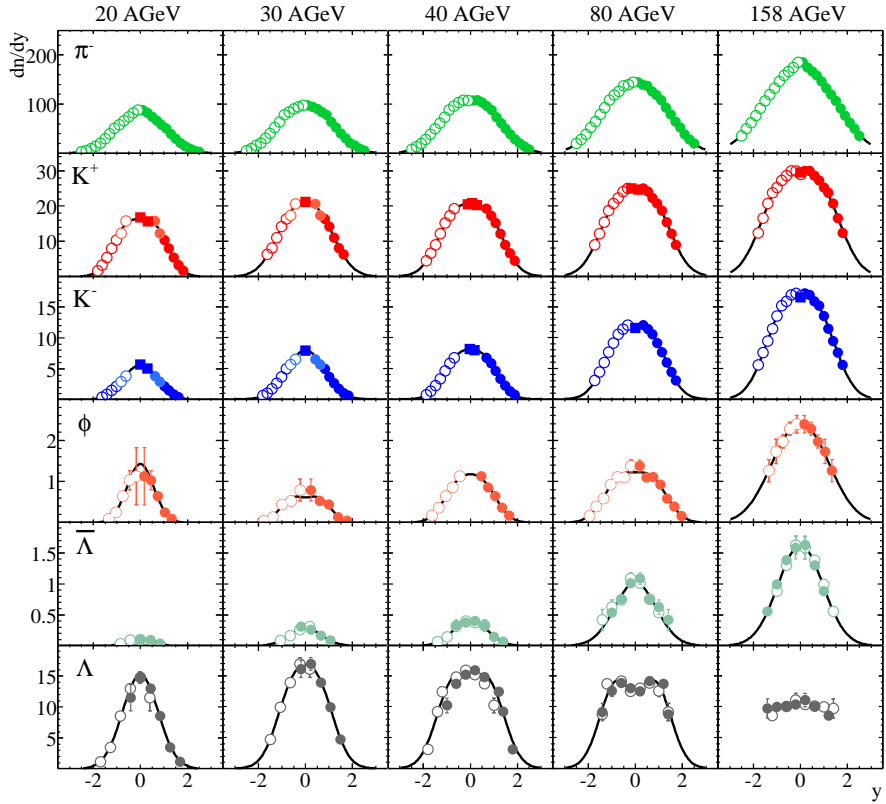,width=12.0cm}
\end{center}
\caption{Rapidity distributions in central Pb+Pb collisions at 20$A$~--~158$A$ GeV. 
Solid dots show measurements, open dots were obtained by reflection around midrapidity.
}
\label{rapall}
\end{figure}

\section{Hadron freezeout parameters from particle yields}\label{chemfo}

\begin{figure}[hbt]
\mbox{
 \parbox{7.0cm}{
  \epsfig{figure=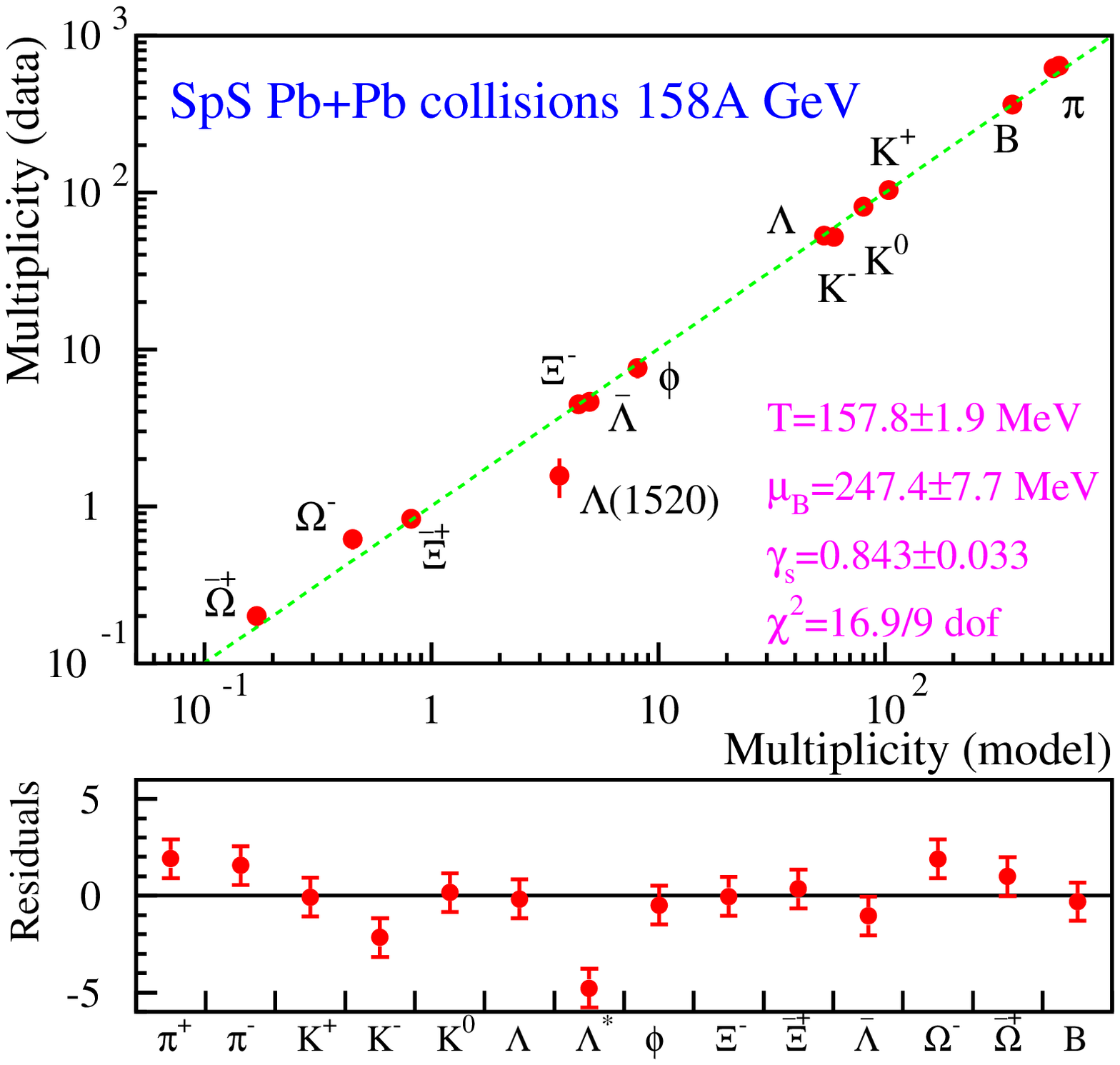,width=7.0cm}}
 \parbox{5.0cm}{
  \epsfig{figure=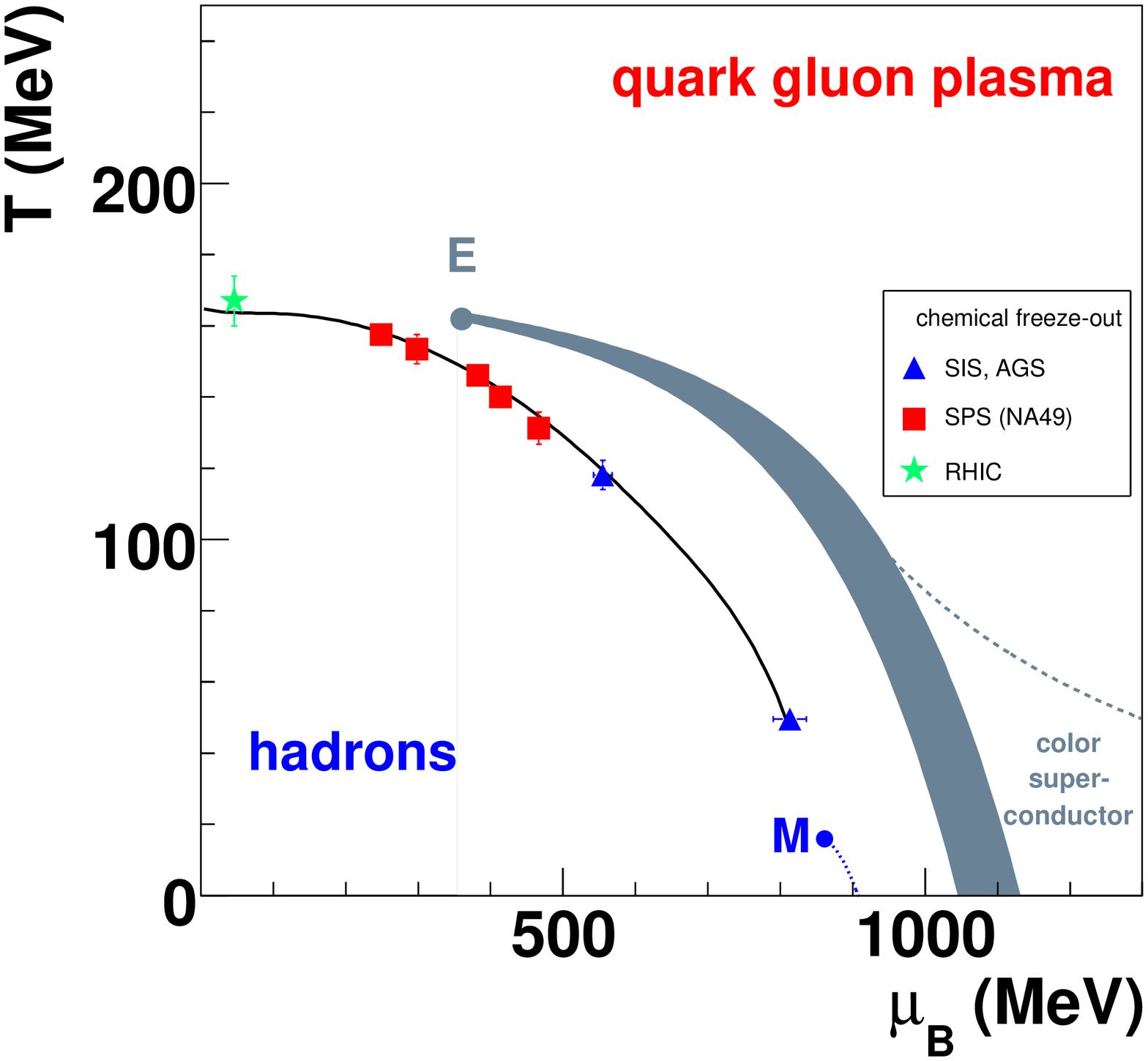,angle=-90,width=5.0cm}}
}
\caption{Left: Results of a statistical model fit to NA49 hadron yields in 158$A$ GeV
central Pb + Pb collisions \cite{beca04}.
Right: Phase diagram of hadronic matter showing the hadron composition freezeout points
obtained from statistical model fits \cite{beca04}. The solid curve represents the
empirical freezeout condition 1 GeV/hadron \cite{cley98}. The grey band indicates the
expected first order phase boundary between QGP and hadrons which ends in a critical point E
estimated from lattice QCD \cite{fod04}.
}
\label{stat_phasedg}
\end{figure}

The large acceptance of the NA49 detector allows measurements of rapidity
spectra from midrapidity up to almost beam rapidity (see fig.~\ref{rapall}).
Rapidity distributions of most particles are peaked at midrapidity. Only $\Lambda$ hyperons,
sharing a valence quark with the projectile, show a flattening at higher collision
energy. Due to the reflection symmetry of Pb+Pb collisions
4$\pi$ yields can be determined.  Particle yield ratios in A+A (as well as elementary)
collisions are consistent with statistical model predictions from threshold to the
highest energies using only 3 parameters:
a temperature T, a baryo-chemical potential $\mu_B$
and a strangeness suppression parameter $\gamma_s$. These parameters characterize
the freezeout of particle composition after which only the momentum distributions
further evolve via elastic scattering. An example of the results of the
fit with the statistical model for central Pb + Pb collisions at 158$A$ GeV \cite{beca04}
is shown in fig.~\ref{stat_phasedg}~(left).

The resulting parameters T and $\mu_B$
\cite{beca04} are plotted in the phase diagram of hadronic matter
in fig.~\ref{stat_phasedg}~(right)
together with the phase boundary predicted by lattice QCD \cite{fod04}.
One observes that the freezeout points at SPS energies approach the phase
boundary and and are closest to the critical point at about 60$A$ GeV.

\begin{figure}[hbt]
\begin{center}
\mbox{
 \parbox{4.5cm}{\hspace{-0.5cm}
  \epsfig{figure=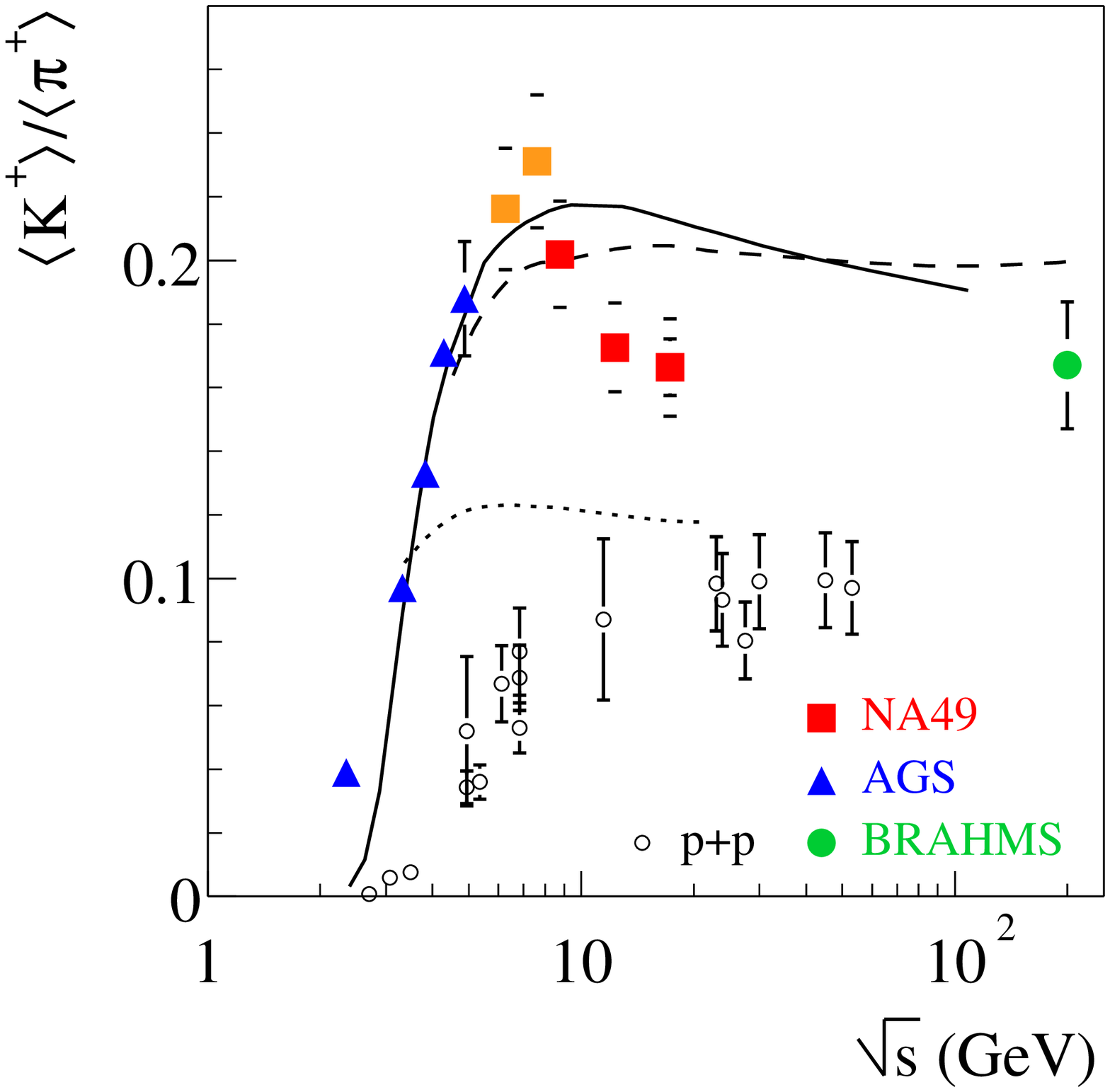,width=4.5cm}}
 \parbox{4.5cm}{\hspace{-0.8cm}
  \epsfig{figure=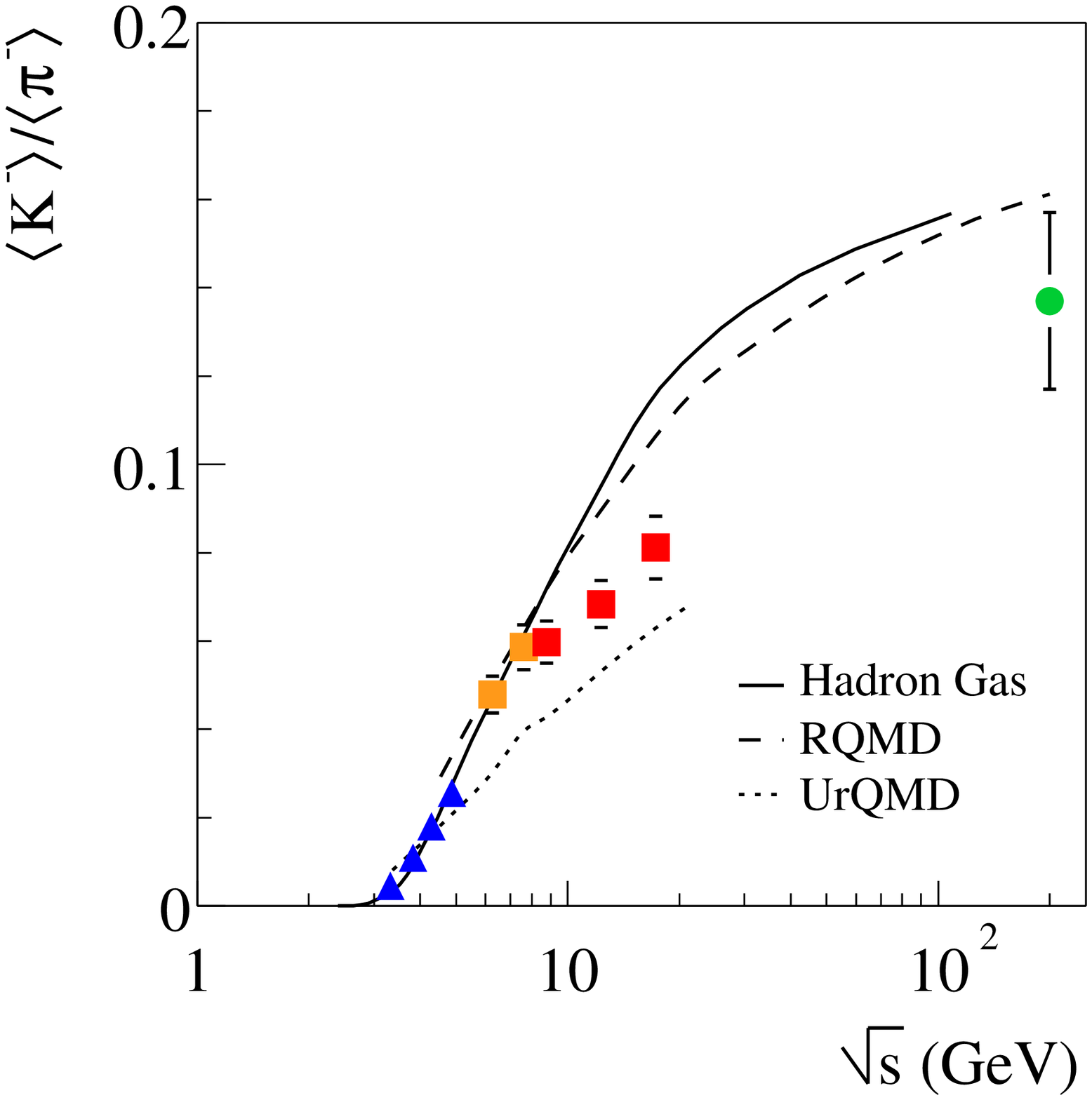,width=4.4cm}}
 \parbox{4.5cm}{\hspace{-1.1cm}
  \epsfig{figure=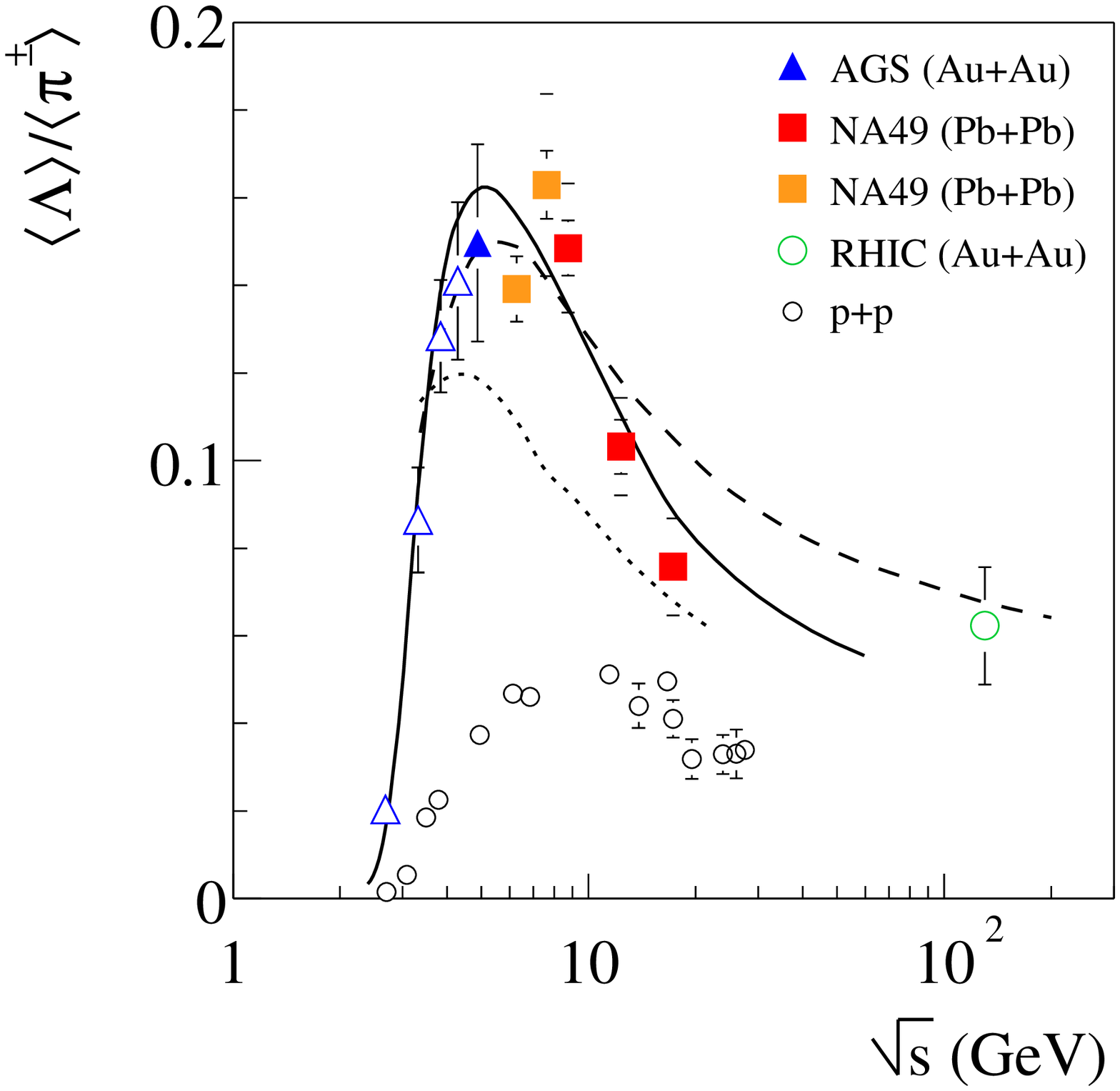,width=4.5cm}}
}
\end{center}
\caption{
Ratio of $4\pi$ yields versus nucleon-nucleon cms energy $\sqrt{s}$:
$\langle K^+ \rangle / \langle \pi^+ \rangle $ (left);
$\langle K^-\rangle / \langle \pi^-\rangle $ (center);
$\langle \Lambda \rangle / (0.5 \cdot (\langle \pi^+ \rangle + \langle \pi^- \rangle))$ (right).
Results from p+p collisions (open dots) and model predictions for Pb+Pb collisions
(curves) are also shown.
}
\label{fig3}
\end{figure}

\section{Indications for the onset of deconfinement}\label{deconf}

A detailed overview of the energy dependence of strangeness production is presented
in fig.~\ref{fig3}. The $\langle$K$^+\rangle$/$\langle\pi^+\rangle$ ratio (left)
shows a steep rise from the threshold of kaon production, a maximum around the 
lowest SPS energy and a decrease
to a somewhat lower plateau value. Although the microscopic transport models RQMD,
UrQMD \cite{rqmd00} and the statistical hadron gas model (full
equilibrium version ($\gamma_s = 1$) supplemented by the
freezeout condition 1~GeV/hadron \cite{cley98}) follow the gross trend, they do not
reproduce the pronounced peak of the $\langle$K$^+\rangle$/$\langle\pi^+\rangle$ ratio at the SPS. 
Since anti-baryon production yields are small, $\langle$K$^+\rangle$ counts essentially
half of all $\bar{s}$ quarks in the final state (the other half is contained in K$^0$).
In contrast $s$ quarks are distributed between anti-kaons and hyperons (mainly $\Lambda$) because 
of the large net baryon density at SPS energies. As a consequence of the rapid decrease of
the net baryon density over the SPS energy range, the sharp peak in
$\langle$K$^+\rangle$/$\langle\pi^+\rangle$ is reflected in a break in the
energy dependence of the $\langle$K$^-\rangle$/$\langle\pi^-\rangle$ ratio (fig.~\ref{fig3}~(center)) 
and a wider maximum
in the $\langle \Lambda \rangle / \langle \pi \rangle$ ratio (fig.~\ref{fig3}~(right)).

\begin{figure}[hbt]
\begin{center}
\mbox{
 \parbox{6.0cm}{
  \epsfig{figure=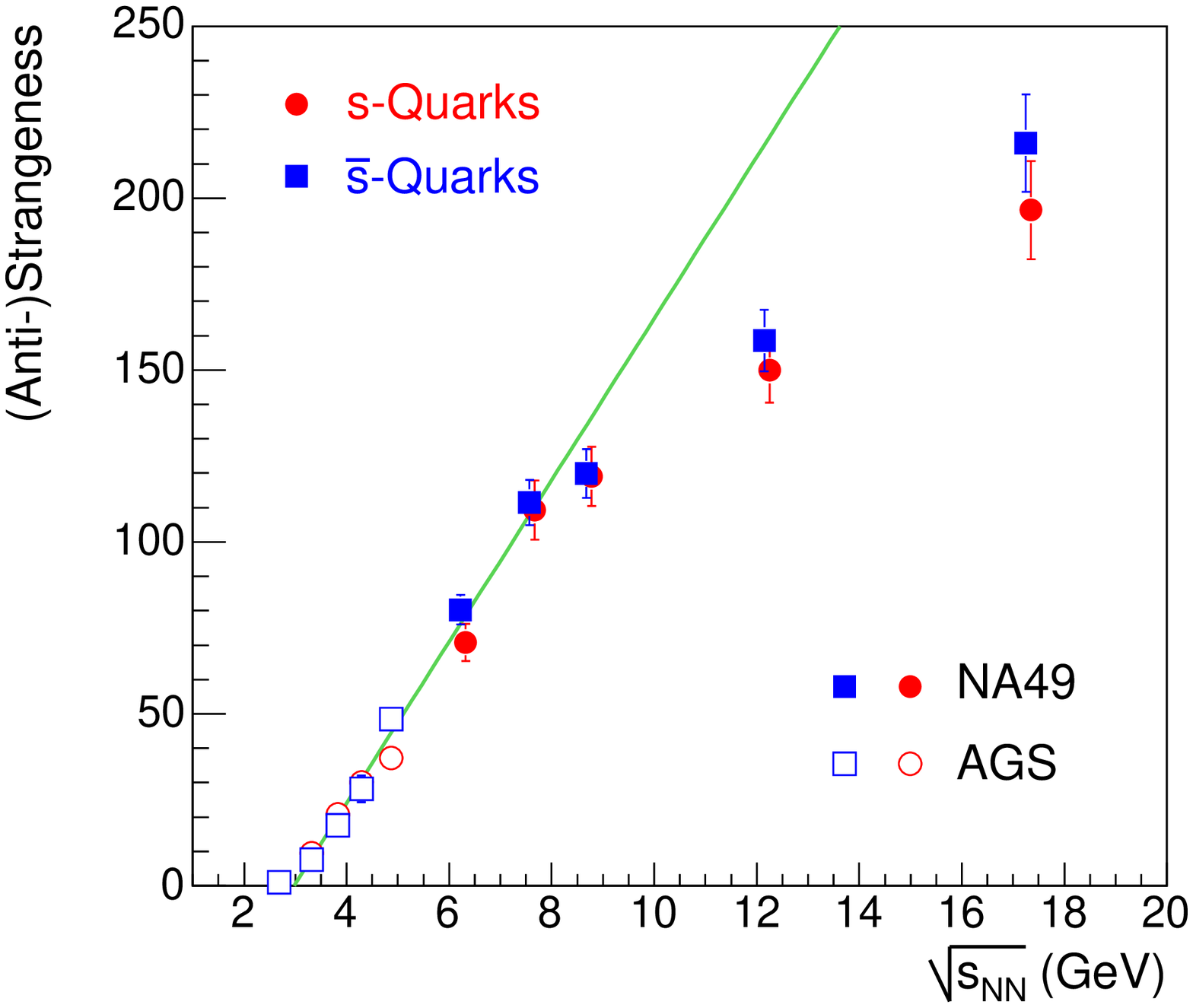,width=6.0cm}}
 \parbox{6.0cm}{
  \epsfig{figure=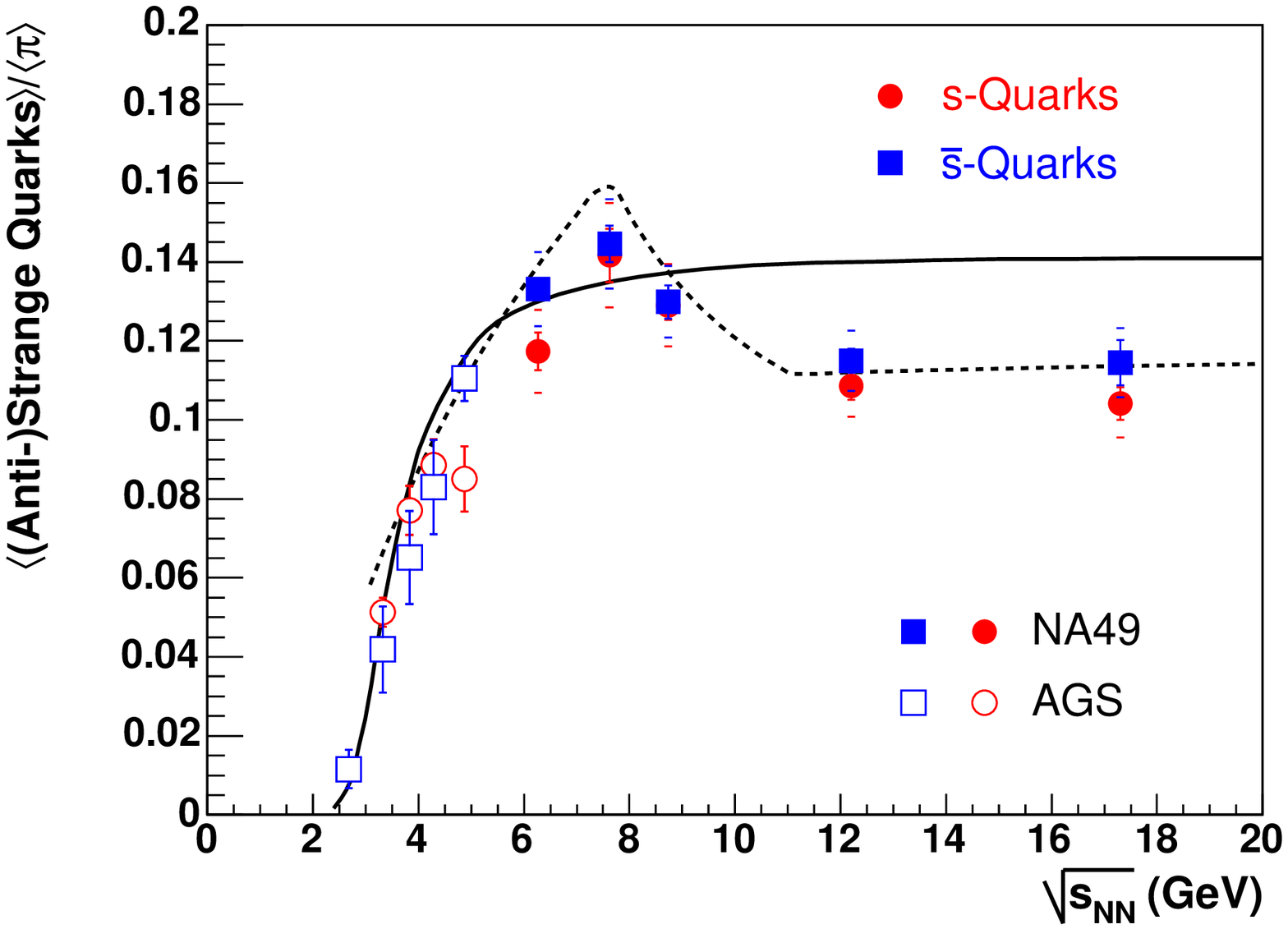,width=6.0cm}}
}
\end{center}
\caption{
Left: Estimated multiplicities of all $s$ and $\bar{s}$ valence quarks contained in
produced particles as a function of energy. The straight line was obtained from a fit to
the low energy points. Right: Ratio of this quantity to the pion multiplicity. 
The full curve shows the prediction of the statistical model using a parameterization
of $\mu_B(\sqrt{s})$ and $\gamma_s = 1$ \cite{cley98}. The dashed curve depicts the
prediction of the SMES model \cite{GaGo99}.
}
\label{fig4}
\end{figure}

Using isospin symmetry as well as predictions from the hadron gas model for some
unmeasured strange particle species one can obtain an estimate of the number
of $s$ and $\bar{s}$ valence quarks contained in the produced particles.
The result shown in fig.~\ref{fig4}~(left) demonstrates that there is a break in the
energy dependence of total strangeness production at 30$A$ GeV. Dividing by the pion
multiplicity (fig.~\ref{fig4}~(right)) confirms the sharp peak already seen in the
$\langle$K$^+\rangle$/$\langle\pi^+\rangle$ ratio. It has been pointed out that
these features coincide with a transition from a baryon to a meson dominated
final state \cite{cley05}. However, as seen from the statistical model curve in
fig.~\ref{fig4}~(right) this effect cannot adequately describe the measurements.

\begin{figure}[hbt]
\begin{center}
\mbox{
 \parbox{4.5cm}{\hspace{-0.5cm}
  \epsfig{figure=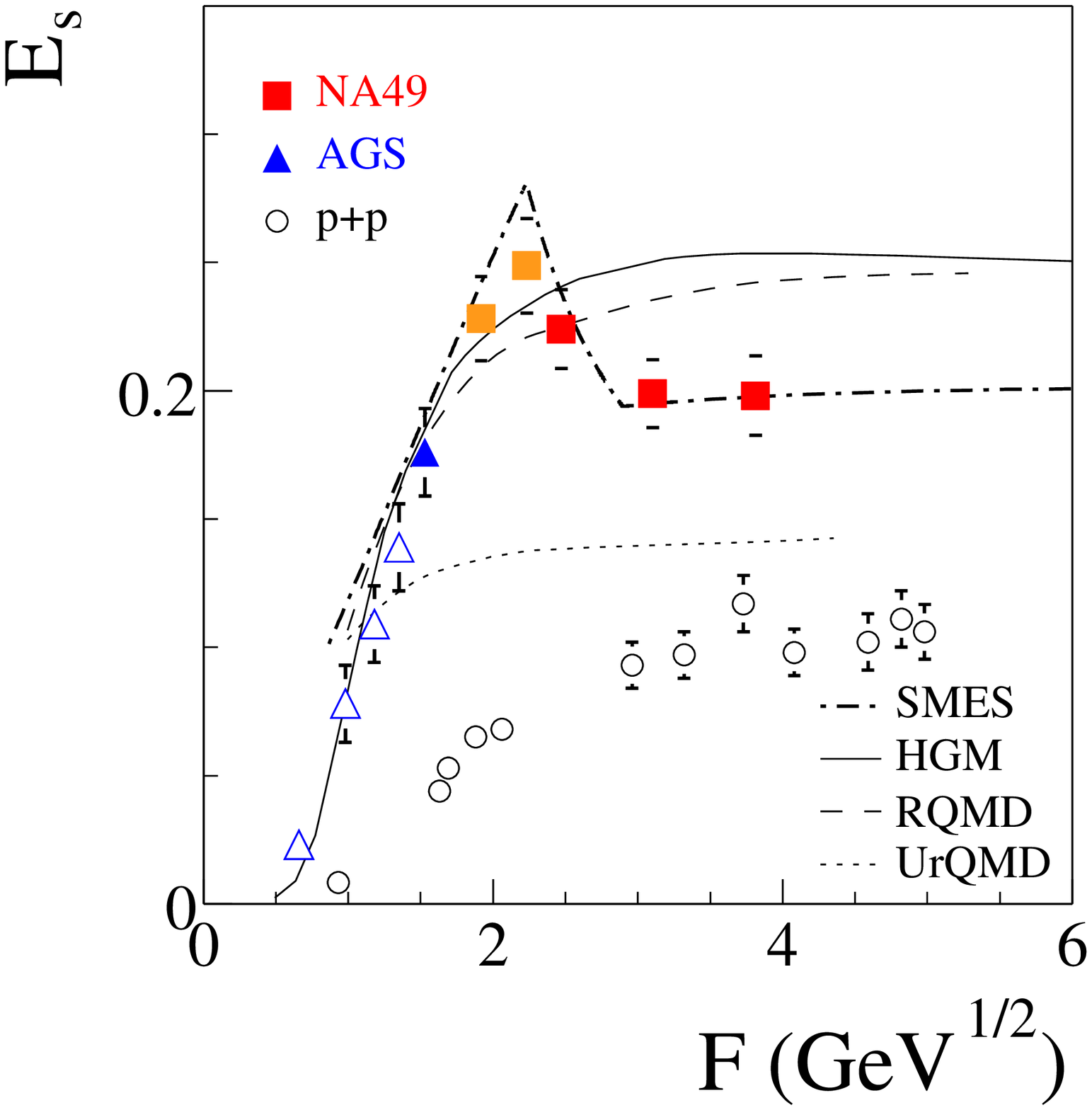,width=4.5cm}}
 \parbox{4.5cm}{\hspace{-0.8cm}
  \epsfig{figure=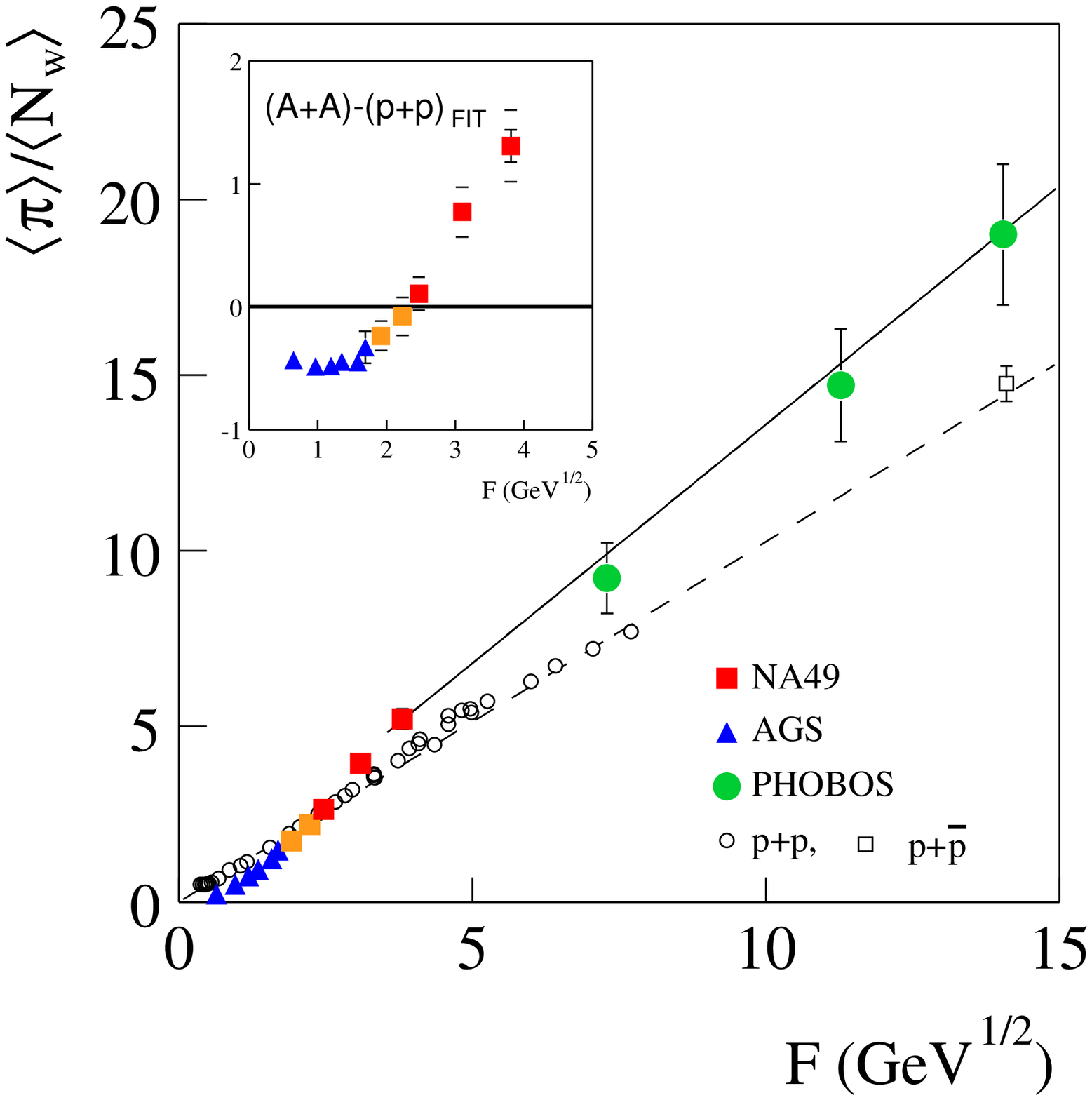,width=4.5cm}}
 \parbox{4.5cm}{\hspace{-1.1cm}
  \epsfig{figure=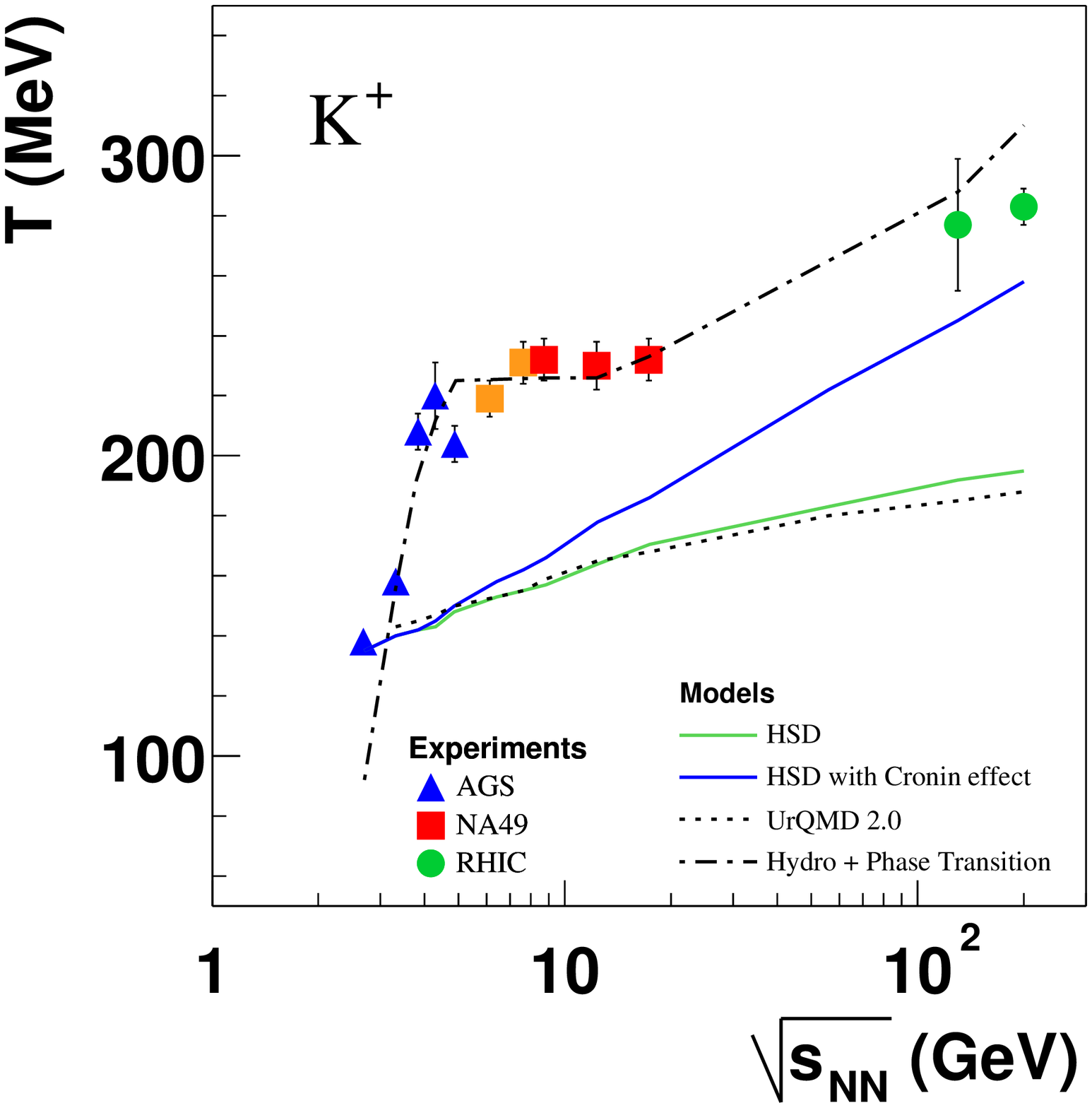,width=4.5cm}}
}
\end{center}
\caption{
Left: Relative strangeness yields
E$_S = (\langle \rm K + \bar{\rm K} \rangle + \langle \Lambda \rangle) / \langle \pi \rangle$
versus Fermi energy variable F$\approx s_{NN}^{0.25}$. Center: Pion yield
per participating nucleon $\langle\pi\rangle$/$\langle$N$_W \rangle$ versus F.
Right: Energy dependence of the effective temperature parameter T of K$^+$
transverse mass spectra. Results from p+p collisions are shown as open
circles, model predictions for Pb+Pb collisions as curves.
}
\label{fig5}
\end{figure}

A close approximation of the total strangeness to pion production
ratio can also be calculated from measured yields via the observable
E$_S = (\langle \rm K + \bar{\rm K} \rangle + \langle \Lambda \rangle) / \langle \pi \rangle$
which is plotted in fig.~\ref{fig5}~(left) as a function of collision energy. As expected,
it shows the same sharp peak followed by a plateau as the
$\langle$K$^+\rangle$/$\langle\pi^+\rangle$ ratio, which is
not seen in p+p collisions and not reproduced by hadronic models. On the other hand this
feature can be understood in a reaction scenario with the onset of deconfinement
around 30$A$ GeV as proposed in the statistical model of the early stage
(SMES \cite{GaGo99}, dash-dotted curve in fig.~\ref{fig5}~(left)).
E$_S$ reflects the behavior of the ratio of the number of s + $\bar{\text{s}}$ 
quarks to entropy in the model.
It shows a steep threshold rise while the system stays in the
hadron phase but drops to the value expected in the QGP at higher collision energies when the
fireball initially reaches the deconfined partonic phase. At the same transition energy of
30$A$ GeV the rate of increase of the number of produced pions per participating nucleon
(a measure of the entropy per baryon in the model) was predicted to increase. As
seen in fig.~\ref{fig5}~(center), this seems to be confirmed by the measurements for collisions
of heavy nuclei, while there is no change in energy dependence for p+p reactions.

\begin{figure}[hbt]
\begin{center}
\epsfig{file=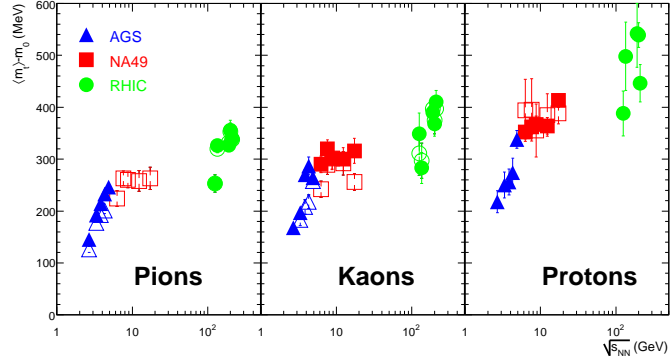,height=5.0cm}
\end{center}
\caption{Energy dependence of $\langle m_t \rangle - m_0 $ for $\pi^-$ and $\pi^+$ (left), 
K$^+$ and K$^-$ (center),and  p and $\bar{\text{p}}$ (right) from AGS, SPS and RHIC experiments.
Full (open) symbols show results for negatively (positively) charged particles.}
\label{mean_mt}
\end{figure}

The transverse mass spectrum of kaons is well described by an exponential
function $e^{ - m_t/T}$. As shown in fig.~\ref{fig5}~(right) for K$^+$ the inverse slope
parameter T rises steeply at low energies, stays at an approximately constant
value through the SPS energy range and ends up slighly higher at RHIC energy.
Similar behavior is seen for K$^-$ (not shown), but is not observed in p+p collisions. 
These features have been attributed
\cite{gore03} to the constant pressure and temperature when a mixed phase is present
in the early stage of the reaction. In fact, a hydrodynamic calculation \cite{hama04}
modeling both the deconfined and the hadronic phases can provide a quantitative
description (dash-dotted curve in fig.~\ref{fig5}~(right)). The spectra of other abundantly
produced particles are not well fitted by exponential functions (see e.g. fig.~\ref{fig1}) but
their shape can be characterised by the average value of the transverse mass  $\langle m_t\rangle$
minus the rest mass m$_0$. The results in fig.~\ref{mean_mt} show that the near constancy 
of the slope of the spectra in the SPS energy range is a common feature for all these particles
species.

\section{Results on fluctuations}\label{fluct}

If the fireball freezeout occurs close to the boundary of a first order phase transition
or near the QCD critical point then
large event-to-event fluctuations may be expected. The NA49 detector was therefore designed
with a large acceptance to allow for a wide range of fluctuation and correlation measurements.

\begin{figure}[hbt]
\begin{center}
\mbox{
 \parbox{6.0cm}{
  \epsfig{figure=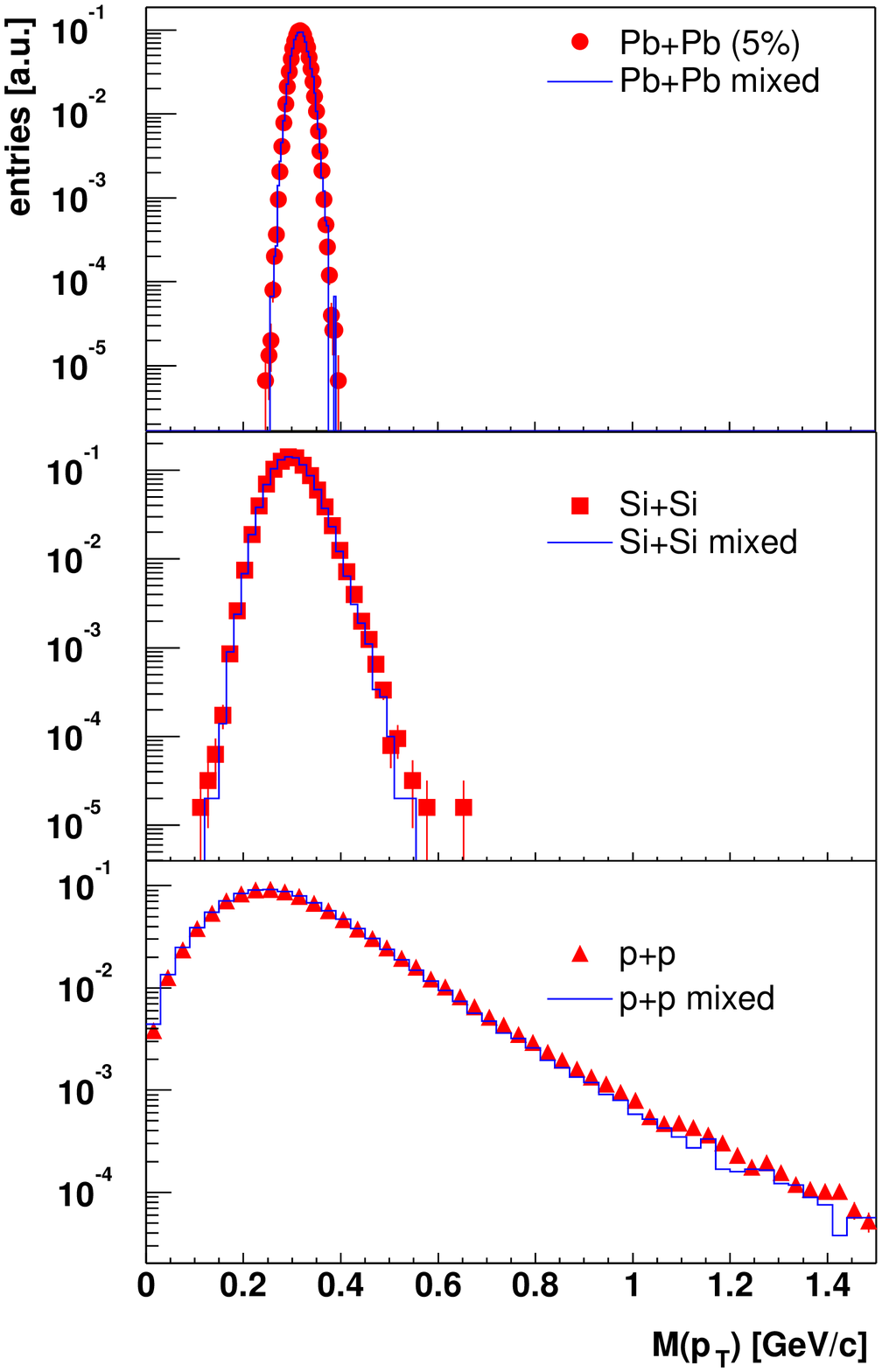,width=6.0cm}}
 \parbox{6.0cm}{
  \epsfig{figure=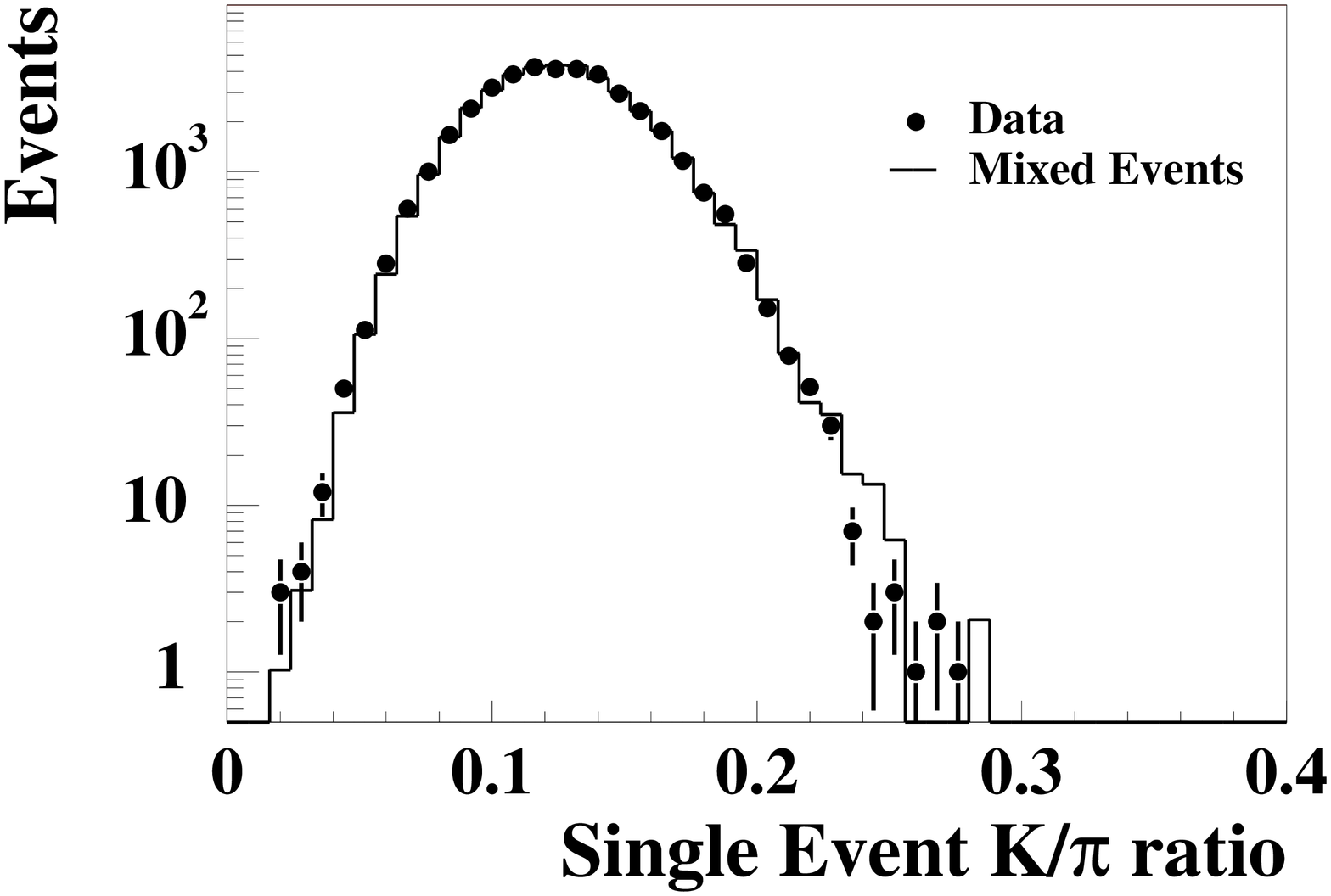,angle=-90,width=6.0cm}}
}
\end{center}
\caption{
Left: Distributions of event-wise $\langle p_T \rangle$ in the rapidity interval
$1.1 < y < 2.6$ for Pb+Pb, Si+Si and p+p collisions at 158$A$ GeV.
Right: Distribution of the event-wise K/$\pi$ ratio in central Pb+Pb collisions
at 158$A$ GeV for laboratory momenta $>$ 3 GeV/c. Results from real events are shown by data 
points from mixed events by histograms.
}
\label{avpt}
\end{figure}

\begin{figure}[hbt]
\begin{center}
 \parbox{6.0cm}{
  \epsfig{figure=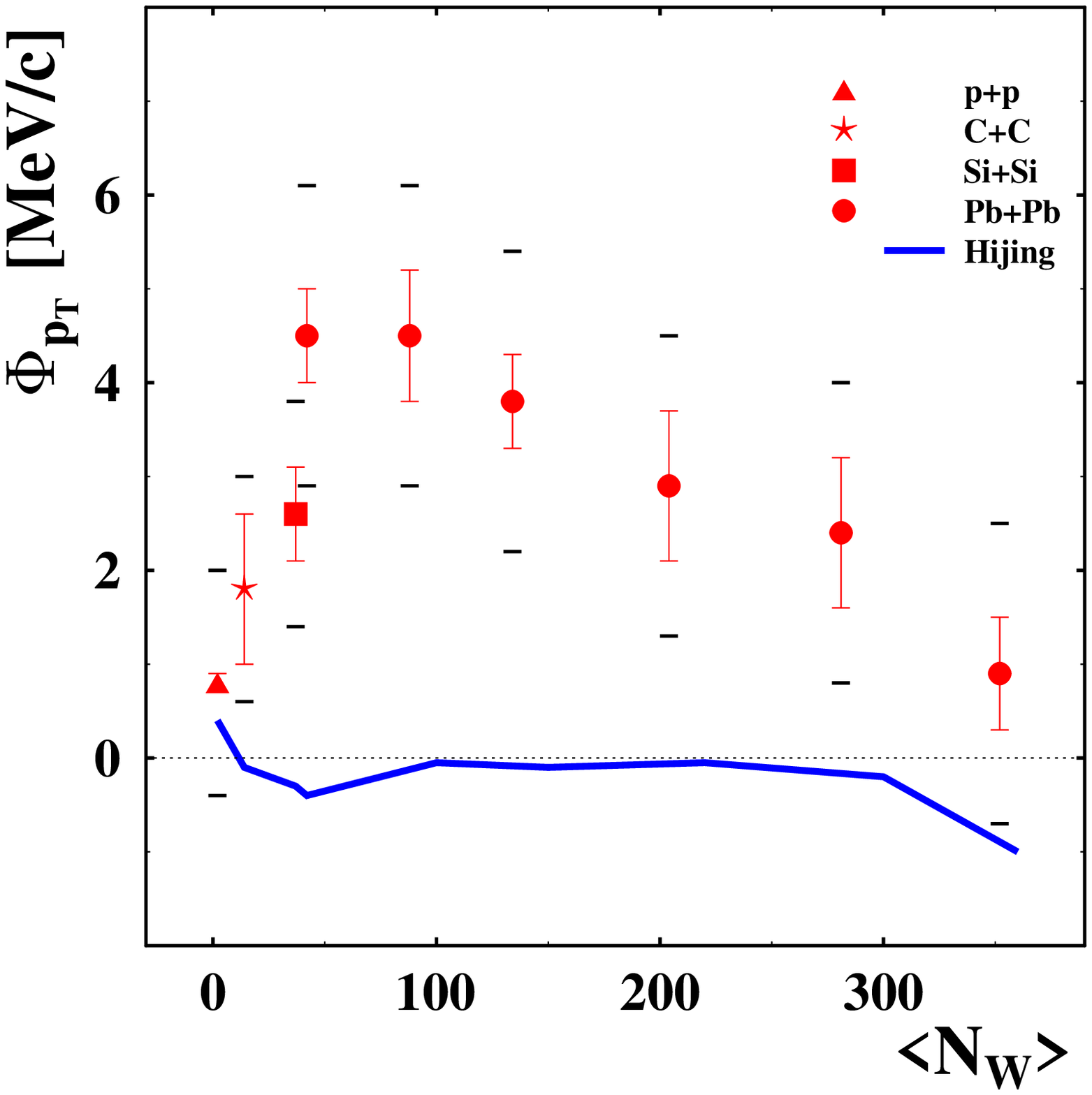,width=6.0cm}}
 \parbox{6.0cm}{\vspace{0.6cm}
  \epsfig{file=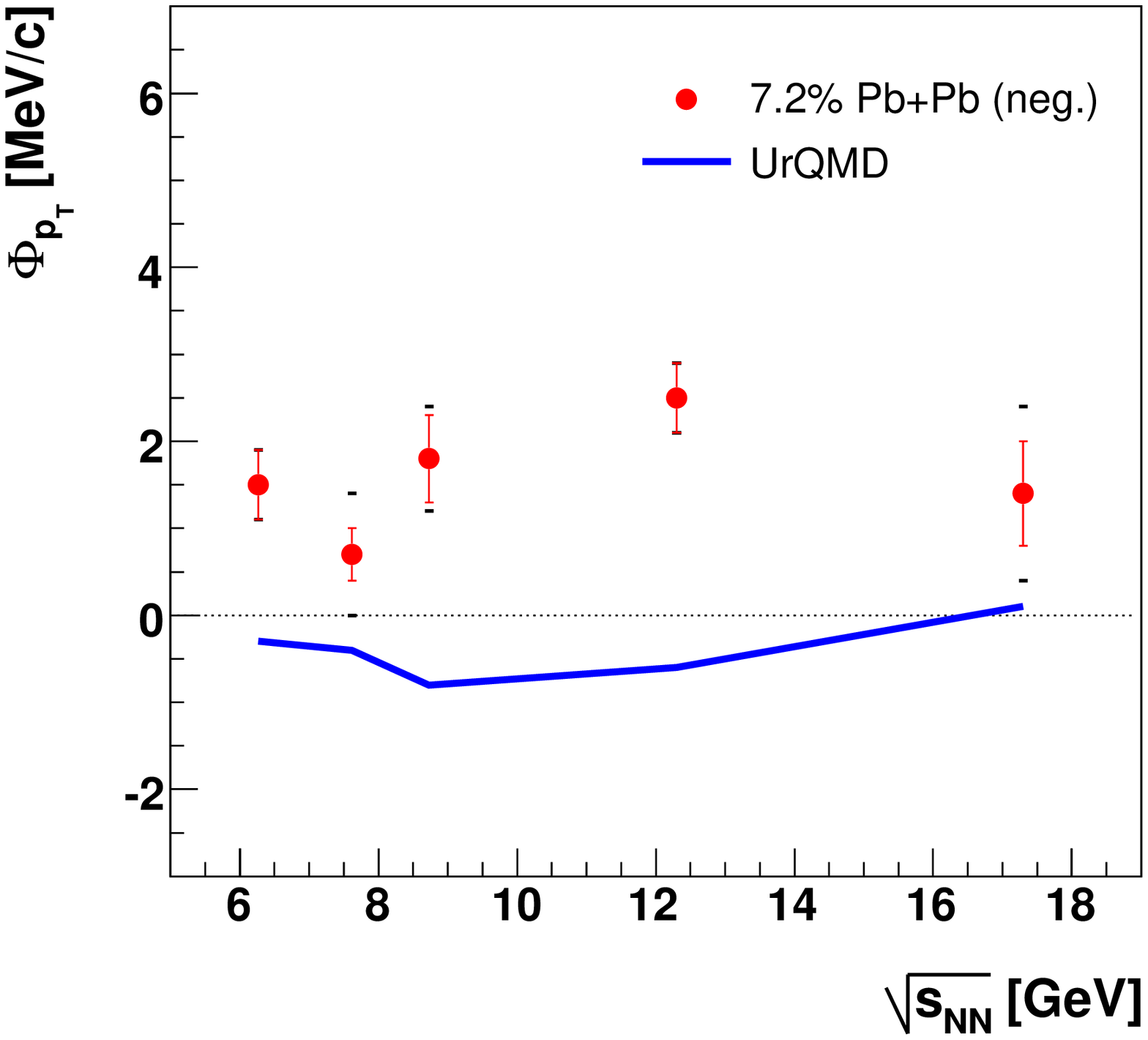,height=6.0cm}}
\end{center}
\caption{
Left: The $\langle p_T \rangle$ fluctuation measure $\Phi_{p_T}$ for
negatively charged particles versus the number of nucleon participants N$_{\text W}$ 
in p+p, C+C, Si+Si and Pb+Pb collisions at 158$A$ GeV. The curve shows predictions 
from the HIJING model filtered through the acceptance of the NA49 detector.
Right (preliminary): Energy dependence of $\langle p_T \rangle$ fluctuation measure $\Phi_{p_T}$ for 
negatively charged particles in central Pb+Pb collisions. The curve shows predictions from 
the UrQMD model filtered through the acceptance of the NA49 detector. 
Results are for the rapidity interval $1.1 < y_{\pi} < 2.6$.
}
\label{avpt_edep}
\end{figure}

The average transverse momentum $\langle p_T \rangle$ in the event is related to the
temperature and radial flow in the fireball. The latter might fluctuate with the
initial energy density and the fraction of deconfined matter at the early stage
of the collision. The distributions for event-wise $\langle p_T \rangle$ are shown
in fig.~\ref{avpt}~(left) for central Pb+Pb , Si+Si and p+p collisions at 158$A$ GeV
by the data points \cite{avgpt}. These are compared to results from mixed events (histograms)
in which correlations are destroyed by construction while preserving the inclusive $p_T$ and
multiplicity distributions. Clearly, the fluctuations are close to the statistical
limit as represented by the mixed events and no distinct event classes are observed.
A quantitative analysis was performed using the quantity $\Phi_{p_T}$ which is defined
to vanish for purely statistical fluctuations and which is furthermore independent of
the number of collison participants in superposition models. The results for $\Phi_{p_T}$ 
in the rapidity range $1.1 < y_{\pi} < 2.6$ are plotted
versus collision system size in fig.~\ref{avpt_edep}~(left). The measured $\Phi_{p_T}$ values are
smaller than 5 MeV/c and thus below a few percent of the $\langle p_T \rangle$. Interestingly
a significant system size dependence is observed with a maximum for peripheral Pb+Pb
collisions. A similar dependence was found for the fluctuations of the multiplicity
of negatively charged particles \cite{na49_qm2004}. The two observations may be related
due to a possible correlation between $\langle p_T \rangle$ and multiplicity \cite{mrow04}.

The energy dependence of the fluctuation measure $\Phi_{p_T}$ at the SPS for
central Pb+Pb collisions is plotted in
fig.~\ref{avpt_edep} (right). The observed values are small and approximately constant. Somewhat
larger results have been obtained near midrapidity by experiment NA45 \cite{adamova_2003}. 
At RHIC the $\langle p_T \rangle$ fluctuations increase substantially due to the rise of jet 
production \cite{adams_2005}.

Fluctuations of the K/$\pi$ ratio are believed to increase near the critical point.
It has also been pointed out \cite{goga04} that at the onset of deconfinement there
could be a significant decrease.
The majority of strange quarks in the produced particles are contained in kaons
at SPS and RHIC energies. The mass of strangeness carriers in the hadron phase (kaons)
is much larger than the temperature whereas in the deconfined phase (strange quarks) 
it is smaller. Therefore fluctuations in the initial energy density at fixed
collision energy are expected to lead to large fluctuations  of the K/$\pi$ ratio
when the system stays in the hadronic phase. However, these K/$\pi$ ratio fluctuations
should significantly decrease if deconfinement occurs during the early stage of
fireball evolution  \cite{GaGo99,goga04}. The SMES model curve in fig.~\ref{fig5} (left)
provides an illustration of this generic prediction when one looks at the change of the
K/$\pi$ ratio for small variations of the energy variable F in the respective regions.

\begin{figure}[hbt]
\mbox{
 \parbox{6.0cm}{
  \epsfig{figure=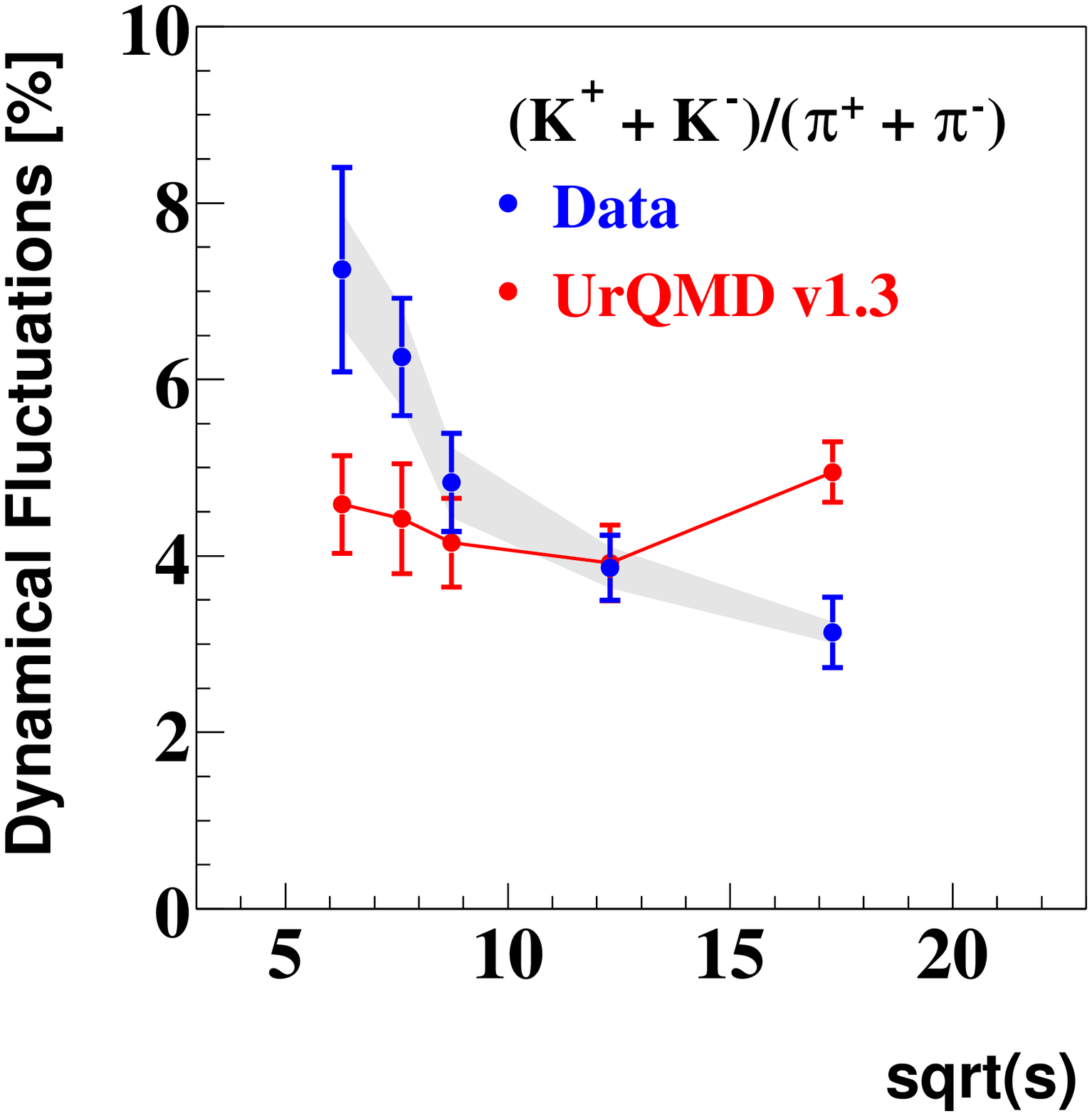,width=6.0cm}}
 \hspace*{0.5cm}
 \parbox{6.0cm}{
  \epsfig{figure=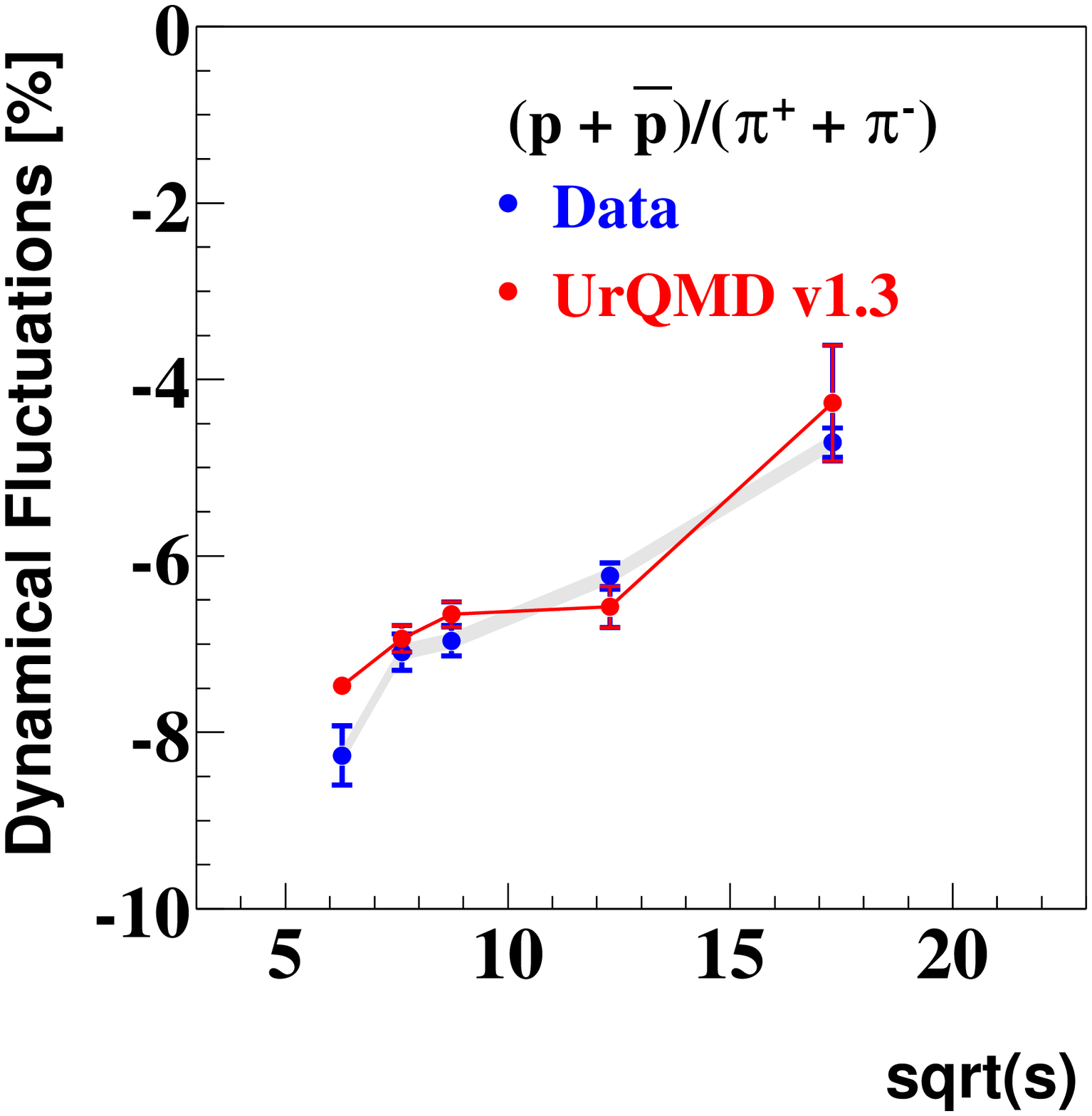,width=6.0cm}}
}
\caption{
Energy dependence of dynamical fluctuations of the $(K^++K^-)/(\pi^++\pi^-)$ ratio (left) and
the $(p + \bar{p})/(\pi^++\pi^-)$ ratio (right) in central Pb+Pb collisions at the SPS.
(preliminary)
}
\label{kpifluct}
\end{figure}

The $(K^++K^-)/(\pi^++\pi^-)$ (as well as $(p + \bar{p})/(\pi^++\pi^-)$) ratio 
was extracted event by event with a maximum 
likelihood fit method for particles with laboratory momenta larger than 3 GeV/c 
in central Pb+Pb collisions at the five SPS energies \cite{rola04}. The momentum
cut is necessary to allow identification by specific energy loss of the particles
in the TPCs. This leads to acceptances which shift to larger cms rapidities with
increasing particle mass. An example of the distribution of the $(K^++K^-)/(\pi^++\pi^-)$ ratio at
158$A$ GeV is shown in fig.~\ref{avpt}~(right) where it is also compared to the
results from mixed events (histogram) \cite{rola99}. Clearly, both distributions
are very similar and indicate that non-statistical fluctuations are small.
Quantitatively, dynamical fluctuations were
calculated by comparing the width $\sigma_{data}$ of the ratio distribution to the one
obtained for mixed events from the formula
$\sigma_{dyn}^2=\sigma^2_{data}-\sigma^2_{mixed}$. As seen from
fig.~\ref{kpifluct}~(left) the dynamical fluctuations of the $(K^++K^-)/(\pi^++\pi^-)$ ratio
amount to only around 5 \% of the average ratio (about 0.15), but they show a decrease 
towards higher energy. Preliminary results from the STAR experiment at RHIC
for higher energies also indicate values close to those at the highest SPS energy \cite{das05}.
Such a behaviour is not seen
in typical reaction models, e.g. UrQMD, but is consistent with the qualitative
expectations from the onset of deconfinement \cite{goga04}. 

The dynamical fluctuations of the $(p + \bar{p})/(\pi^++\pi^-)$ ratio are plotted
for the five SPS energies in fig.~\ref{kpifluct}~(right). The observed negative values
are described quantitatively by the UrQMD model and are most likely due to the correlated
production of p and $\pi$ via baryon resonances.

\begin{figure}[hbt]
\mbox{
 \parbox{6.0cm}{
  \epsfig{figure=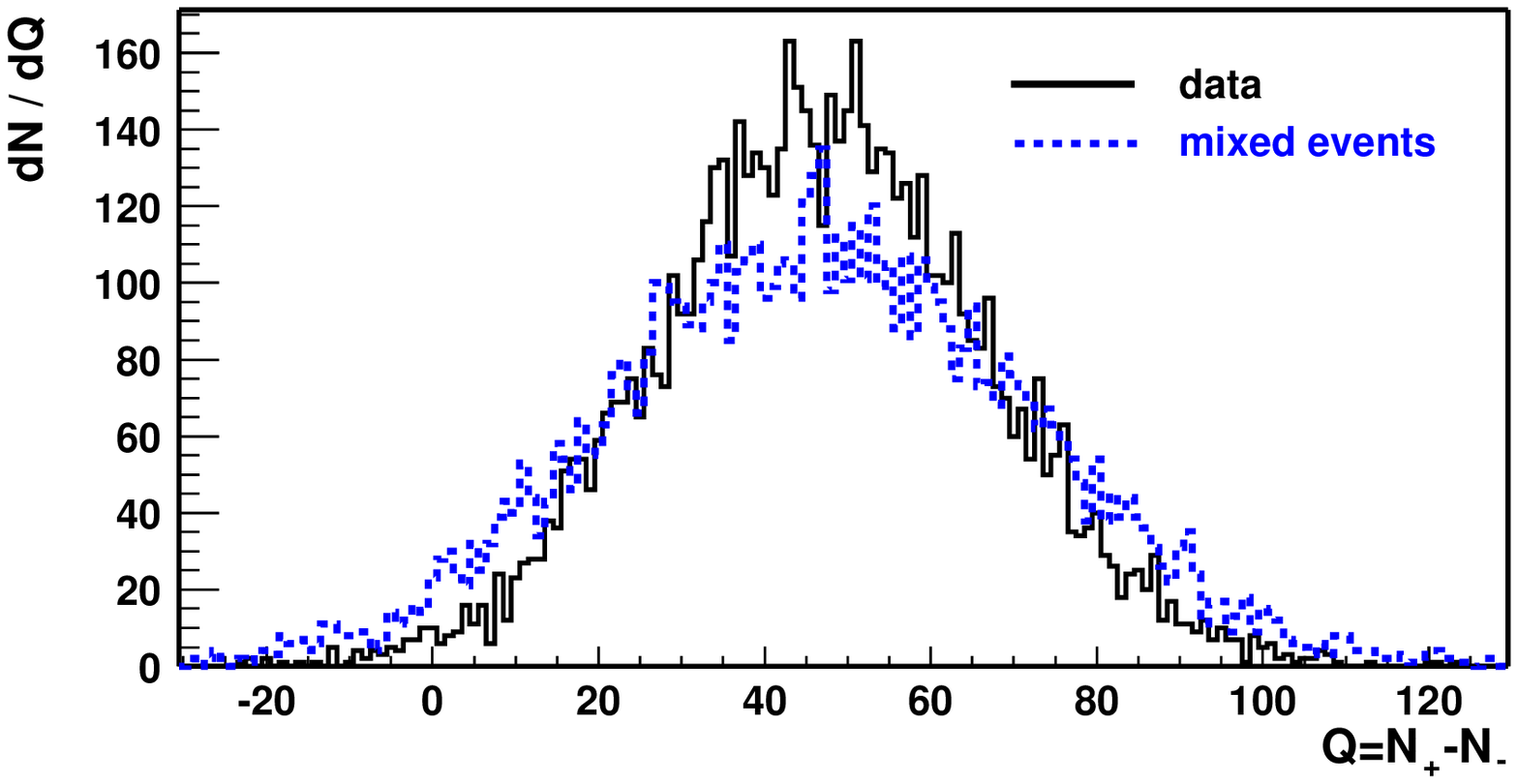,width=6.0cm}}
 \parbox{6.0cm}{
  \epsfig{figure=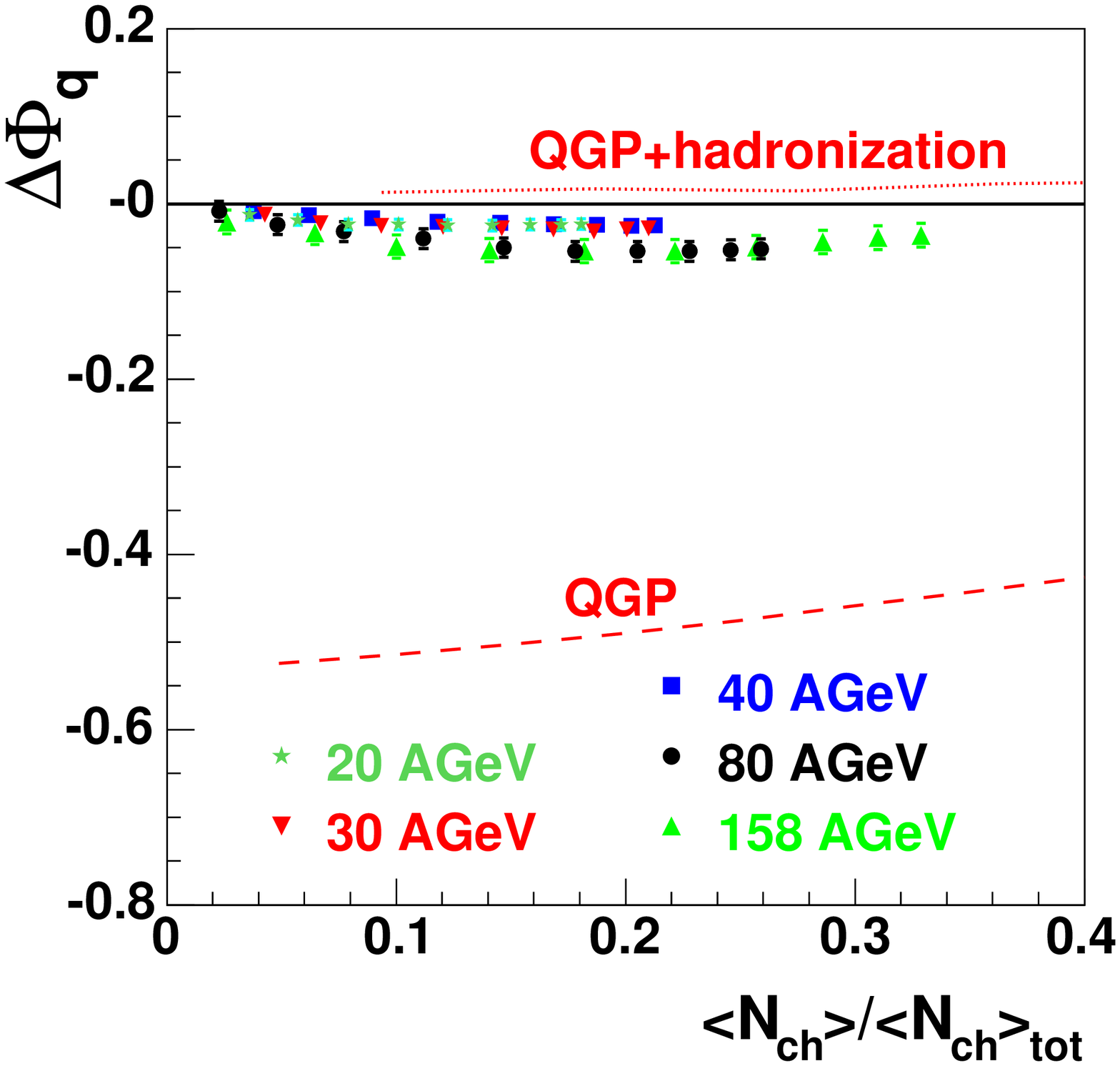,height=6.0cm}}
}
\caption{
Left: Distibutions of net charge Q = N$^+$ - N$^-$ for the papidity range $-1.4 < y < 2.4$
in central Pb+Pb collisions at 158$A$ GeV for data (full histogram) and mixed events
(dotted histogram).
Right: Measure $\Delta\Phi_{\text q}$ of electric charge fluctuations corrected for charge
conservation versus rapidity acceptance.
}
\label{chflct}
\end{figure}

Electric charge fluctuations were predicted to be strongly reduced if the
fireball passed through a QGP phase \cite{jeon00}. Fig.~\ref{chflct}~(left) depicts
the distribution of net charge in the papidity range $-1.4 < y < 2.4$ for
central Pb+Pb collisions at 158$A$ GeV as an example. As a consequence of charge
conservation the data (full histogram) display a narrower distribution than
mixed events in which this effect is removed.  Fig.~\ref{chflct}~(right) shows the
measurements of electric charge fluctuations in terms of
the variable $\Delta\Phi_q$ \cite{alt04} in which effects of global
charge conservation are subtracted. Neither significant energy dependence nor
the predicted reduction are observed. However, more recent model calculations suggest
that hadronisation effects \cite{bial02} and at SPS also resonance decays \cite{zar02}
probably mask the reduction.

\begin{figure}[hbt]
\begin{center}
\mbox{
 \parbox{6.0cm}{\hspace{-0.5cm}
  \epsfig{figure=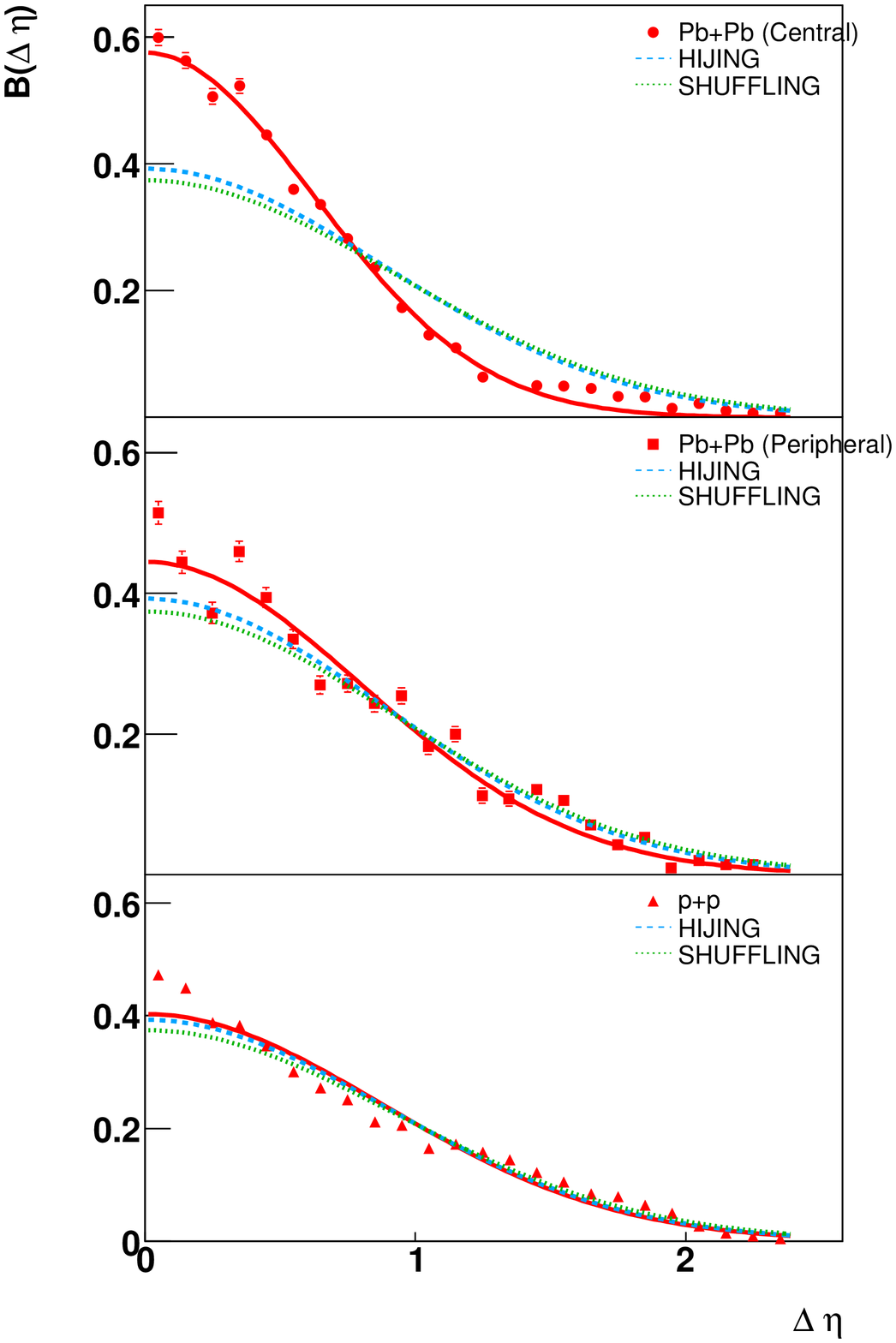,width=6.0cm}}
 \parbox{6.0cm}{
  \vspace{-6.0cm} \hspace{-0.5cm} 
  \epsfig{figure=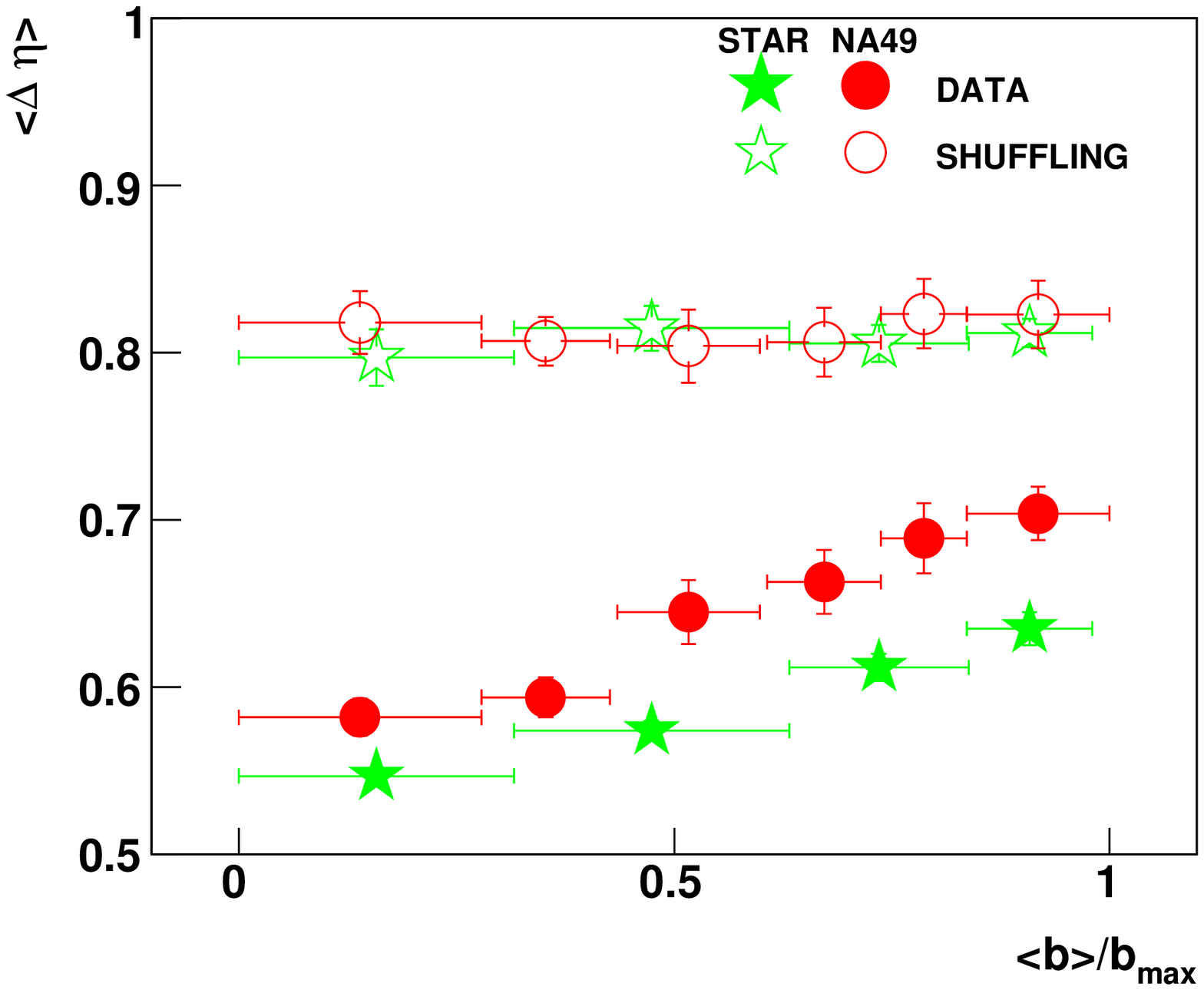,height=5.0cm}}
 \parbox{6.0cm}{
  \vspace{4.5cm} \hspace{-6.5cm}
  \epsfig{figure=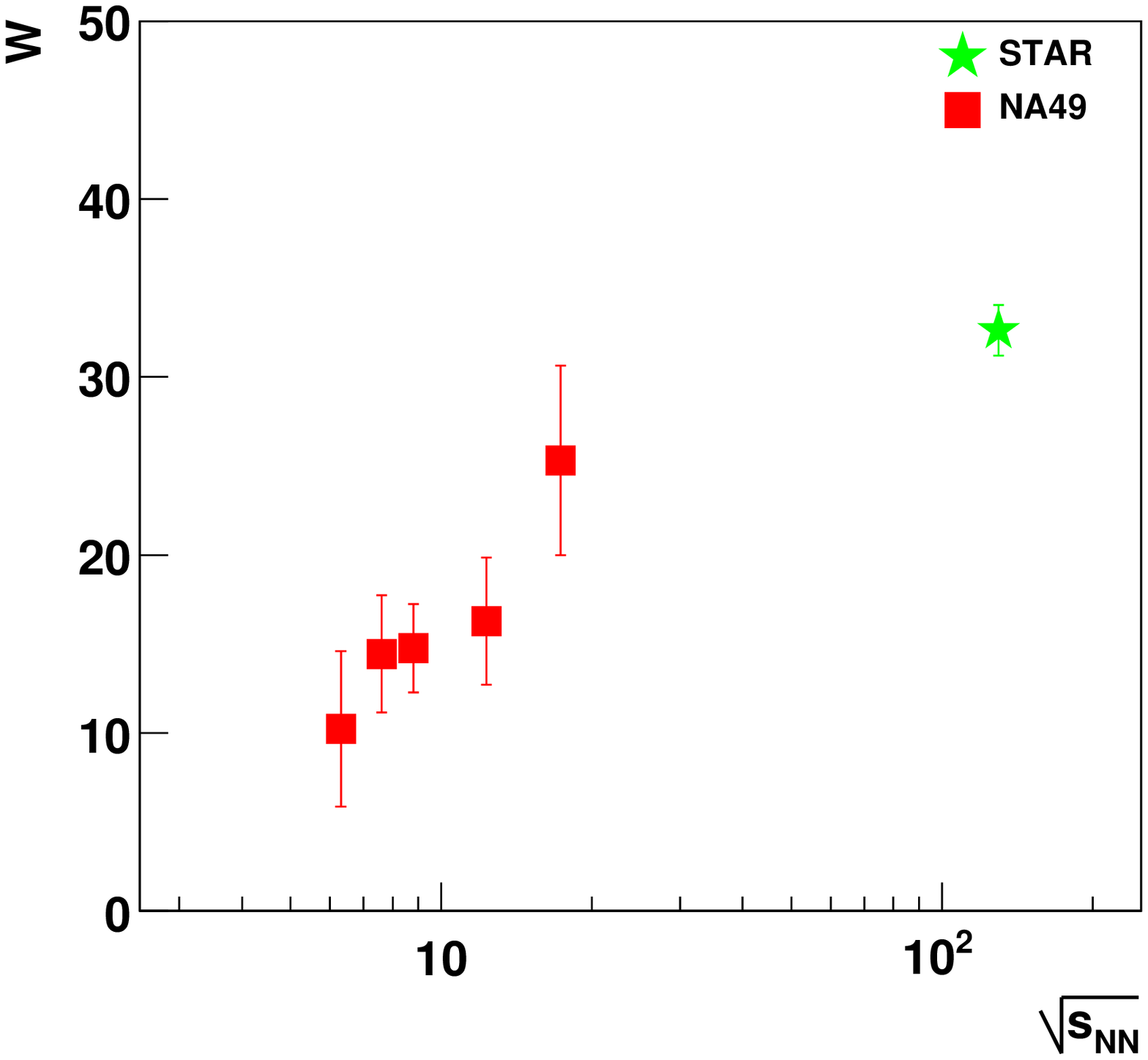,height=5.5cm}}
}
\end{center}
\caption{
Left: Balance function versus pseudo-rapidity difference of oppositely charged particle
pairs at midrapidity and top SPS energy. Curves show Gaussian fits to data (solid), 
shuffled events (dotted) and HIJING simulation (dashed).
Right,~top: Width of the balance function versus impact parameter at midrapidity in  Pb+Pb
collisions from NA49 at SPS (dots) and STAR at RHIC (stars). Solid symbols denote data,
open symbols show results from shuffled events.
Right,~bottom: Energy dependence of the normalized narrowing parameter W of the balance function. 
}
\label{balance}
\end{figure}

A first order phase transition from the deconfined to the hadronic phase during the
early stage of Pb+Pb collisions is expected to result in a softening of
the equation of state and consequently in a long lifetime of the mixed phase. 
The study of the balance function (BF) was suggested as a method to investigate
the time of hadron formation \cite{bass00}. In many models pions are assumed
to be created in charge neutral pairs. Due to the buildup of longitudinal flow
and rescattering during the evolution of the fireball, the rapidity correlation
of opposite charges will be diluted for pairs formed early in the process but will
remain narrow for late hadronisation as expected in a mixed phase scenario. This
correlation is quantified by the BF which is plotted as a function of the
pseudorapidity difference $\Delta\eta$ of the pairs in fig.~\ref{balance}~(left). The
data are compared to shuffled events, in which the pseudo rapidities of the
particles are scrambled while the charges are retained. In this way one obtains
an estimate of the maximum decorrelation under the condition of charge conservation.
It is evident that the data show increasingly shorter range opposite charge
correlations beyond the effects of charge conservation as the centrality of 
Pb+Pb collisions increases.

The widths $\langle \Delta\eta \rangle$ are calculated numerically 
from the BF distributions and the results
are plotted in fig.~\ref{balance}~(right,top) for the top SPS energy \cite{bal05} and 
for RHIC \cite{adams03}. The width decreases by 17 \% respectively 14 \% for central compared
to peripheral collisions for data while it remains constant for shuffled events. Thus the
predicted narrowing is observed, although it is now believed that a large part of this
effect is due to the increasing strength of radial flow. In fact a model
assuming hadron formation by quark coalescence as well as collective radial flow can
reproduce the observed width of the BF \cite{bial04}.

In order to be able to
compare measurements at different energies and with different acceptance one may calculate
the normalized difference (in percent) 
W~=~$100 \cdot (\langle\Delta\eta^{\text shuff}\rangle - \langle\Delta\eta\rangle)$)/$\langle\Delta\eta^{\text shuff}\rangle$ 
of the widths of the BF for shuffled and real events. The energy dependence of the 
normalised narrowing parameter W is plotted in fig.~\ref{balance}~(right,bottom). 
It shows a steady rise, i.e. more narrowing of the BF with respect to uncorrelated 
particle production throughout the SPS energy range up to RHIC.  One does not observe a
significant structure at 30$A$ GeV, where other observables suggest the onset of deconfinement.

\section{Conclusions}\label{concl}

Results from a comprehensive study of nucleus--nucleus collisions in the SPS
energy range were presented. The produced matter shows strong
transverse and longitudinal flow. Ratios of yields of produced particles are
approximately consistent with statistical equilibration. The freezeout parameters
of the hadron composition are close to the phase boundary predicted by QCD.
The main charcteristics of nucleus--nucleus collisions at the highest SPS energy are 
quite similar to those found at RHIC.
  
The energy scan of central Pb+Pb collisions revealed that at around 30$A$~GeV the
ratio of strangeness to pion production shows a sharp maximum, the rate of increase
of the produced pion multiplicity per wounded nucleon increases and the effective
temperature of pions and kaons levels to a constant value.  These features are not reproduced 
by present hadronic models but find a natural explanation in a reaction scenario with the onset
of deconfinement in the early stage of the reaction at low SPS energy.
Further support for this scenario is provided by the decrease of fluctuations of
the K/$\pi$ ratio in the SPS energy range. On the other hand, measurements of fluctuations 
of the average transverse momentum, of the electric charge, as well as of
correlations such as the balance function and $\pi\pi$ Bose-Einstein correlations exhibit 
a smooth energy dependence. 

The present results suggest that further study of the role of the initial volume in the onset of
deconfinement and the search for the predicted critical point of QCD are of
great interest. This led to an initiative for a low energy program at RHIC \cite{rhic_lowe}
and the proposal of a future light-ion program at the SPS \cite{eoi,loi}.

\section{Acknowledgement}\label{acknow}

My sincere thanks to the organizers of the XLVI School of Theoretical Physics in
Zakopane 2006 for supporting my attendance and inviting me to present a review of
results from the NA49 experiment. I would also like to express my gratitude to
all members of the NA49 collaboration for their hard and diligent work that produced
the comprehensive results presented in this lecture.

\end{document}